# Any Old Tom, Dick or Harry:
# The Citation Impact of Author First Name Genderedness

Maxime Holmberg Sainte-Marie[*] and Vincent Larivière[**]

[*]mhsm@sdu.dk
ORCID: 0000-0002-0760-0290
Department of Design, Media and Education Science, Southern University of Denmark, Denmark
UNESCO Research Chair on Open Science, École de Bibliothéconomie et des Sciences de l'Information, Université de Montréal, Canada
Chaire de recherche du Québec sur la découvrabilité des contenus scientifiques en français, Université du Québec à Montréal, Canada

[**] vincent.lariviere@umontreal.ca
ORCID : 0000-0002-2733-0689
UNESCO Research Chair on Open Science, École de Bibliothéconomie et des Sciences de l'Information, Université de Montréal, Canada
Chaire de recherche du Québec sur la découvrabilité des contenus scientifiques en français, Université du Québec à Montréal, Canada
Consortium Érudit, Université de Montréal, Canada
Observatoire des Sciences et des Technologies, Centre Interuniversitaire de Recherche sur la Science et la Technologie, Université du Québec à Montréal, Canada

This paper examines the relationship between the genderedness of authors' first names and citation distributions in US-affiliated scholarly production from 2010 to 2019. Merging a first name genderedness table derived from Wikidata with Web of Science publication data, we develop a relative distributional framework that compares name, article, and citation distributions along a continuous genderedness spectrum. The lexical structure of the corpus proves stable across author roles, while productivity diverges substantially by disciplinary group: physical sciences show a consistent masculine lean, social sciences a feminine one, with amplitude remaining stable across groups despite these directional differences. Citation analyses net of publication volume reveal a pervasive deficit across disciplinary groups and author roles: only unambiguously masculine names accumulate citations in excess of their publication volume, while feminine, neutral, and moderately masculine names all fall short of it. This asymmetry is strongest in the life sciences and weakest in the physical sciences except among middle authors, where the pattern reflects the gender composition of large collaborative teams rather than evaluative bias in citing behavior. Taken together, these results are consistent with the hypothesis that first name genderedness shapes citation recognition through implicit bias operating in low-deliberation evaluative contexts.

## 1 Introduction

During the late 1960s and 1970s, initiatives in biomedical and feminist research contributed to the emergence of a clear scientific distinction between sex and gender: the former, together with maleness and femaleness, concerns biologically driven characteristics, whereas gender and its related attributes of masculinity and femininity concern self-identity and culturally determined behavioral and psychological traits (Torgrimson & Minson, 2015; West & Zimmerman, 1987; Reisner et al., 2015; Ansara & Hegarty, 2014). More generally, the concept of gender came to refer to all the subjective aspects surrounding the concreteness and objectivity of sexually determined features. Despite this clear conceptual divide, however, research often reduces gender to sex across disciplinary groups and countries. In social science for example, participants are often asked to disclose their gender by choosing between only two possible answers, namely those historically and conceptually associated with sex: 'male' and 'female' (Westbrook & Saperstein, 2015). Regardless of the intent, this dichotomization of a concept "not at all limited to two kinds" (Walker and Cook, 1998, pp. 255-256) risks obscuring "what the researchers are aiming at and what participants respond to - bodily attributes, legal, gender, or self-defined gender identity" (Lindqvist et al., 2021, p. 334).

In science studies, while a great deal of research effort has been aimed at highlighting gender disparities within the scientific community (Leahy, 2006; Fox, 2001; Xie & Shauman, 2003; Sugimoto & Larivière, 2023), most of it is based on the use of name-to-gender inferencing: automatic procedures whose goal is to determine, from the first name of individuals, whether they are male or female (Gonzalez-Salmón et al., 2024; Halevi, 2019; Mihaljević et al., 2019). From a functional standpoint, these procedures are similar to what the biological and forensic sciences respectively call sexing and sex estimation (Biederman and Shiffrar, 1987; Krishan et al., 2016): in each case, a binary male or female category is inferred by applying learned rules and heuristics to observable features used as proxies, from the form of a fledgling and the frame of a skeleton to the first name of an author. While these procedures are conventionally described as inferring gender and have yielded valuable evidence of gender disparities in science, the category they actually return is the binary male/female of sex, and this operational dichotomization overlooks what precisely distinguishes the concept from that of sex: its grounding in social identity, cultural inscription, and individual self-representation rather than in anatomical or chromosomal traits. This conflation is, partly, a contingency of English: Walker and Cook (1998) note that the language is unusual in having let its grammatical gender system erode to the point where gender became available as a polite synonym for sex, a substitution far less available in languages whose grammatical gender system remains robust enough to keep the term anchored to its linguistic sense (Walker and Cook, 1998).

The first obvious consequence of this dichotomization of gender is that it ignores the fact that many individuals do not fit or identify with either category. When performed systematically by algorithmic tools (Hamidi et al., 2018), this binary classification can have material consequences for non-binary individuals who are misrepresented or rendered invisible by the resulting data (Mihaljević et al., 2019). Another dimension of scholarly gender that is obscured in this process is social identity: in many activities of crucial importance to the scholarly community, the biological maleness or femaleness of individuals might not matter as much as how they are "read by others" as being male or female (Butler, 1990; Litwiller, 2020; Keller, 1999; Bornstein, 1994; Mackay, 2021). The latter scenario, strongly reminiscent of the idea that "masculinity and femininity do not exist 'out there' in the world of objective realities… [but] only in the mind of the perceiver" (Bem 1987, p. 309), finds perhaps its most direct and intuitive application in the analysis of citation behavior.

The act of citing is, at its core, a decision made by scholars about whose work to acknowledge, engage with, and lend visibility to. Neither neutral nor purely intellectual, citation behavior is often shaped by rhetorical, social, and partly non-scientific considerations (Gilbert, 1977; Bornmann and Daniel, 2008; MacRoberts and MacRoberts, 1989). In the typical case, and especially for authors unknown to the citer, citation decisions are largely free from formal oversight and are rarely made explicit or rationalized. These conditions of low deliberation are precisely those under which implicit bias is known to operate most freely (Greenwald and Banaji, 1995). This susceptibility is especially consequential given the role citations play as the primary currency of the reward system of science (Merton, 1968, 1973): beyond the acknowledgement of an intellectual debt, a citation now carries an exchange value whose cumulative effects have an increasingly decisive bearing on the distribution of recognition, funding, and influence, both within and beyond the scholarly ecosystem (Sugimoto and Larivière, 2018; Rossiter, 1993). In this sense, the act of citing is both the scholarly practice most susceptible to implicit bias and the one in which the latter can have the most severe and lasting impact.

In the specific case of gender, the impact of this dual function of citations operates primarily through a single channel: the first name. For readers unfamiliar with the author, first names are quite often the only bibliographic element carrying gender cues, as evidenced by the predominance of name-to-gender services in bibliometric research (Gonzalez-Salmón et al., 2024; Halevi, 2019; Mihaljević et al., 2019). However, the vast majority of first names do not refer to biological sex in a completely unambiguous way. Seen from a subjectivist or phenomenological perspective, what readers do when they believe an author to be male or female is infer their sex based on their own assessment of the relative masculinity or femininity of the author's first name. In this perspective, what the literature typically attributes to a sexually dichotomised conception of gender may therefore very well stem from the genderedness of authors' first names, a conflation made all the more plausible by the low-deliberation context of citation behavior, given the reasonable assumption that most citing authors are unlikely to personally know those they cite.

It is the conjunction of the subjective foundations of citations as exchange value and the singular and ambiguous nature of the first name as a gender cue that motivates the current research question: "to what extent is first name genderedness associated with citation distributions?". Following a short overview on the psychological and social aspects of the relationship between names and gender, we then describe the process behind the construction and merging of the first name genderedness and bibliometric databases used in this project, followed by an analysis and discussion of the results obtained.

# 2 Research Context

Naming is a sociolinguistic universal used by all human societies to distinguish individuals (Bright, 2003): every human being gets assigned a personal name, grows up embodying it, and learns to integrate a community and interact with others on that basis. In shaping how individuals perceive themselves and are perceived by others, names act as powerful cultural signals and as such can carry a lot of information about their bearers as well as the societies in which they live. Research indeed shows that first name features correlate strongly with predicting variables such as income, social class, happiness, education, and occupational prestige, even when controlling for other background factors such as race, gender, parents' education levels, age, and siblings (Aura and Hess, 2004, 2010). Amongst the various signals names can give to oneself and to others, gender certainly figures amongst the most prominent ones (Sue and Telles, 2007). While some languages and cultures use personal names to sex individuals, that is, to explicitly mark whether an individual is male or female, most of them name individuals in a relatively gendered way, depending on how masculine or feminine a given name is to a given culture at a given point in time. In English, first names like 'Jill' and 'Bill' are strongly associated with female and male individuals respectively (Bauer and Coyne 1997; Nick 2017). Because of these strong associations, experimental studies looking at the effect of gendered names can easily choose examples that are overwhelmingly associated with males or females (Lammers et al., 2009; Bengtsson et al., 2012; Foschi and Valenzuela, 2012). However, such clear-cut and experimentally useful associative patterns are far from the cultural norm: some first names like 'Lee' defy easy gender categorization (van Fleet and Atwater 1997), while the genderedness of others can also change over time and country, as is respectively the case with 'Leslie' (Blevins and Mullen, 2015) and 'Andrea' (Vasilescu et al., 2014; West et al., 2013). Gendered naming practices and norms can also be found across cultures: on the one hand, males are more likely to be named after kin and be given historical, biblical, more serious or stable, tradition-bearing names; on the other, feminine names exhibit greater temporal or cross-cultural variability and are more often chosen for aesthetic or expressive reasons (Rossi, 1965; Lieberson, 2000; Hough, 2000). In fact, all first names unfold and move within a vast genderedness spectrum ranging from the most feminine first names to the most masculine ones. Any subtle variation in spelling can result in wide shifts in genderedness: "a woman could have a traditionally male name (such as 'Patrick'), an ambiguous name ('Pat'), or a typically female name ('Patricia'). A continuous spectrum of possibilities ranges between these points; something like 'Patric' might be mostly male but less certainly so than is 'Patrick'" (Urbatsch, 2018, p. 59). And every time a newly born baby is given a name, something changes in the genderedness spectrum.

This spectral conceptualization of gender echoes a broader and still active tradition in the psychometric study of gender, where the masculinity and femininity of individuals are represented as a matter of degree rather of kind, of probabilistically derived quantities rather than of binaries (Del Giudice, 2024; Lippa and Connelly, 1990). As regards to first names themselves, research across disciplines shows that operationalizations focusing on their relative masculinity or femininity allow to draw insights and account for behavioral patterns that a sex-based dichotomy cannot. Part of this evidence relates to what in law and economics has been called the Portia Hypothesis: named in honor of the Shakespearian character who disguised herself as a male lawyer to defend her husband's friend, this hypothesis postulates that individuals with masculine first names, whether male or female, are more likely to be successful in legal professions. This hypothesis was first put to the test by Coffey and McLaughlin (2009), who found that women bearing more masculine first names were significantly more likely to advance to judgeships than otherwise comparable women with less masculine names. In a subsequent study, the authors showed that the masculine-name advantage on lawyers' earnings persisted even after controlling for family wealth and reputation (Coffey and McLaughlin, 2016). Various studies have observed similar Portia effects on hiring and evaluation in musical (Goldin and Rouse, 2000), psychological (Steinpreis, Anders, and Ritzke, 1999), and STEM professions (Moss-Racusin et al., 2012), where applications bearing masculine or male-associated names were consistently rated more favorably than otherwise identical applications bearing feminine ones. The pattern extends to the electoral arena as well: in United States state-legislative elections, candidates with more masculine-sounding names, whether men or women, have been found to receive a larger vote share, implying a corresponding penalty for feminine-sounding names (Urbatsch, 2018).

Beyond the Portia effect, research in developmental psychology shows that boys with female-sounding names tend to misbehave disproportionately upon entry to middle school (Figlio, 2007); in the same vein, a study focusing on pairs of highly-achieving sisters reports that femininely-named girls tend to be as successful as their less femininely-named sisters in mathematics, but are nonetheless less likely to select the most advanced mathematics and science courses in high school (Figlio, 2005). Finally, two science studies projects focused on the impact of perceived gender, based on journal editors' assessments of authors' first names, on publication acceptance rates: the first study reported evidence of acceptance bias against feminine-named submitters in one of the journals examined (Tregenza, 2002); the second discovered that the establishment of a double-blind reviewing process in a primary research journal resulted in increased acceptance rates for authors with feminine first names (Budden et al. 2008), although the magnitude of this effect was subsequently debated (Webb et al., 2008; Engqvist & Frommen, 2008). Taken together, these findings suggest that the perceived masculinity or femininity of a first name can function as an independent signal in evaluative contexts, shaping outcomes in ways that a binary classification of authors by sex would be unable to detect. Regardless of the unaddressed question of the origin of these effects, first name genderedness consistently emerges as a significant predictor of outcomes across domains, providing sufficient motivation for the distributional analyses that follows.

# 3 Methodology

Gender information is typically absent from bibliographic database metadata and must therefore be obtained from external sources. While some studies collect it directly through surveys (Zheng et al., 2022), administrative databases (Larivière et al., 2011), input from colleagues (Amering et al., 2011) or by analyzing pronouns or portraits used on websites and social media (Maliniak et al, 2013; Azoulay and Lynn, 2020; Kong et al., 2022; van den Besselaar & Sandström, 2017), most research on gender disparities in the scholarly community rely on third-party or in-house name-to-gender inference procedures, which match first names against databases that associate names with gender frequencies, labels or probabilities (González-Salmón et al., 2024; Santamaría and Mihaljević, 2018). While some models allow for finer-grained inferences by adjusting for country and year of birth (Boekhout et al., 2021; Costas et al., 2015; Blevins and Mullen, 2015), all existing procedures are however affected by two related construct validity problems. The first is relatively well documented (González-Salmón et al., 2024; Santamaría and Mihaljević, 2018; Halevi, 2019): they reduce multi-valued gender classifications to binary, sex-based categories, producing an operationalization that misrepresents the very construct it purports to measure. The second problem is that, by assigning to each individual author a property (i.e. gender) that the data captures at the level of the first name, these procedures conflate two distinct entities: the carrier of the gender signal (the first name) and its target (the person that bears it). This entity conflation problem has received far less explicit attention: while the social risks of algorithmic gender recognition are increasingly discussed (Hamidi et al., 2018; Keyes, 2008; Lockhart et al., 2018), the resulting mismatch between the unit at which gender is measured and the unit to which it is attributed has rarely been treated as a construct validity problem in its own right.

The present study addresses both construct validity problems. The entity conflation problem arises specifically when a signal carried by the first name is attributed to the individual who bears it; by taking the first name itself as the sole unit of analysis, with citation counts and gender frequencies both aggregated at that level, no property is transferred across entities, and the mismatch does not arise. This dissolves the conflation rather than correcting it, at an explicit and accepted cost: the analysis bears on the genderedness of first names rather than directly on the sex of the authors who bear them. We adopt this narrower object of study deliberately and specify in the Results the conditions under which a claim about name genderedness can still be read, probabilistically, as a claim about author sex. As for the dichotomization problem, it is addressed by operationalizing first name genderedness as a continuous, relative frequency-based variable rather than as a binary label, yielding a measure that more faithfully represents the multi-valued nature of the construct as well as its relation to scholarly production. The remainder of this section describes the data collection and statistical analysis procedures developed to implement this operationalization.

### 3.1 Data Collection and Preprocessing

All gender-related data used in this research has been extracted from semantic triples contained in the Wikidata knowledge base. This choice of database was motivated by four considerations. First, it builds on a procedure previously developed and validated by Bérubé et al. (2020), where Wikidata was used to extract first-name frequency data in a bibliometric context, providing a methodological precedent for the present application. Second, Wikidata is a freely accessible, versioned open knowledge base whose complete data dumps can be easily downloaded, making the analysis fully reproducible and independently verifiable. Third, its entity-level representation of named individuals, linked to biographical metadata, provides direct access to the absolute frequencies underlying any genderedness score: for any given name, the raw counts of male and female entities can be retrieved and used to compute the score from first principles, without relying on pre-computed probability estimates whose derivation would be opaque. Finally, this access to absolute frequencies facilitates sensitivity analyses and methodological variants that would be difficult to implement with data sources that expose only aggregate probability estimates. These advantages come at the cost of a known coverage bias toward Western, English-language, and male biographical entries (Konieczny & Klein, 2018; Tripodi, 2023), the implications of which we address in the Limitations, where we also note that WoS itself carries parallel biases that partly attenuate the asymmetry introduced by the matching step.

Following the download of the truthy (i.e. most failsafe) data dump of April 3rd, 2024, all Wikidata semantic triples having human-specific entities as subjects and given name (P735), male (Q6581097), or female (Q6581072) as objects were extracted from the downloaded data. Following this, first names were converted to ASCII encoding format using the Python Unidecode package (Šolc, 2025) and all entities extracted were merged using semantic triplet subjects as keys, resulting in a total of 65,263 unique first names spread across 7,807,233 human entities. Finally, a gender frequency table was generated by compiling the frequency of male and female entities for each distinct first name. As regards to bibliometric data, author first names, roles, and article citation counts were extracted from the Web of Science (henceforth WoS) on April 30th, 2024. To control for cultural, national, and historical factors, only articles, notes, and reviews published between 2010 and 2019 by authors affiliated to American institutions were collected[1]. Five different subsets were collected, depending on the following publication roles of authors: single author (62,768 unique first names over 386,274 documents), first author (329,670 unique first names over 3,696,564 documents), middle author (621,322 unique first names over 15,246,000 documents), last author (215,252 first names over 3,696,564 documents), and corresponding author (383,429 unique first names over 5,882,797 documents). In cases where corresponding author info was absent, single authors or first authors were also considered corresponding authors by default. Finally, discipline assignation for each article was done using the National Science Foundation (NSF) field classification of journals used in the Science and Engineering Indicators (SEI) reports (National Science Foundation, 2006). By assigning only one of the 14 disciplines to each journal, each article is thus assigned only one discipline, which allows for a complete disciplinary partitioning and avoids any double counting of papers. For analysis and visualization purposes, each of these 14 disciplines were then grouped into 4 disciplinary groups: life sciences, physical sciences, social sciences, and arts & humanities.

[1] We acknowledge that citation counts for articles published toward the end of this window are right-censored relative to earlier articles, as the data extraction of April 2024 allows only five years of accumulation for 2019 publication versus up to fourteen years for 2010 publications. Analyses are therefore best interpreted as reflecting relative distributional patterns rather than absolute citation magnitudes. Because this censoring applies uniformly across the genderedness spectrum within each publication year, it is unlikely to generate the systematic feminine deficit reported here, though we cannot fully exclude an interaction with the disciplinary composition of names.

Before merging both databases, first names in both datasets were then harmonized through a series of preprocessing operations. All non-word characters were replaced with single whitespaces, and substrings of whitespaces were reduced to single whitespaces. To remove initials, all first names were stripped of single letters and substrings of consecutive upper-cased letters (to cover the case of initials without punctuation). All first names were then converted to lower case, and data points which had their first name information deleted because of the previous operations were removed from both datasets. Following these harmonization procedures, gender frequencies from Wikidata first names were assigned to WoS first names through a three-step matching procedure. The first step consisted in attempting to match every WoS first name to an identical Wikidata entry, which resolved 22,970 of the 422,812 unique first names in the WoS dataset (5.4%). This figure is low by construction rather than by failure. After initials are stripped, the 422,812 unique strings still contain a long tail of compound and multi-component names, rare or non-standard spellings, and transliteration variants, so exact whole-string matching resolves only a small share of distinct types. Those types are nonetheless the common, frequently borne names that dominate the corpus by occurrence, accounting for 71.7% of author occurrences and 70.6% of citations. A token-by-token cumulative procedure was then carried out to resolve unmatched compound and multi-component first names (e.g. "mary jane"): each WoS first name was split into its constituent tokens (e.g., "mary" and "jane"), and the male and female entity counts of each corresponding first name in Wikidata were summed up together, with unmatched tokens ignored. This second matching step resolved 174,641 of the 399,842 remaining first names (43.7%). Finally, the WoS names still unmatched were submitted to a character-level Forward Maximum Matching algorithm (Sproat and Shih, 1990) aimed at handling merged first names (e.g. "maryjane"). For each first name, this iterative algorithm finds the longest prefix of the character string that corresponds to a complete Wikidata entry, compiles the associated frequencies, removes the matched prefix, and repeats this process until no further match can be found. This third matching step resolved 169,609 of the 225,201 remaining names (75.3%), amounting to a total of 367,220 matched WoS first names (86.9%), accounting altogether for 90.3% and 90.9% of all dataset articles and citations respectively. Finally, all remaining unmatched first names and corresponding WoS entries were not considered for this study.

The UpSet plot of Figure 1 shows the distribution of the matched WoS first names across the five author role subsets and their intersections (Lex et al., 2014). Looking at author role set size (horizontal bars), the five author role subsets differ substantially in size: the middle author category covers on its own close to 80% of all first names (79.7%), followed by corresponding (49.3%), first (42%), last (26.2%), and single (8.8%) authors. This prominence most likely reflects the structural logic of multi-author collaboration: middle-author positions draw from a larger and more diverse pool of first names than any other role, partly because large collaborative projects concentrate the overwhelming majority of their contributors there. This reading is consistent with the hyperauthorship practices first documented by Cronin (2001), whereby a single article may list hundreds or thousands of contributors, a trend that began in particle physics and has since spread across fields (Singh Chawla, 2019). This hypothesis is further corroborated by the extreme concentration of names exclusive to the middle position: of the 367,220 distinct first names, 160,076 (43.6%) occur exclusively in the middle author position, accounting for 91.3% of all first names confined to a single role, the only other role with any exclusive names being the last author position (8.7%).

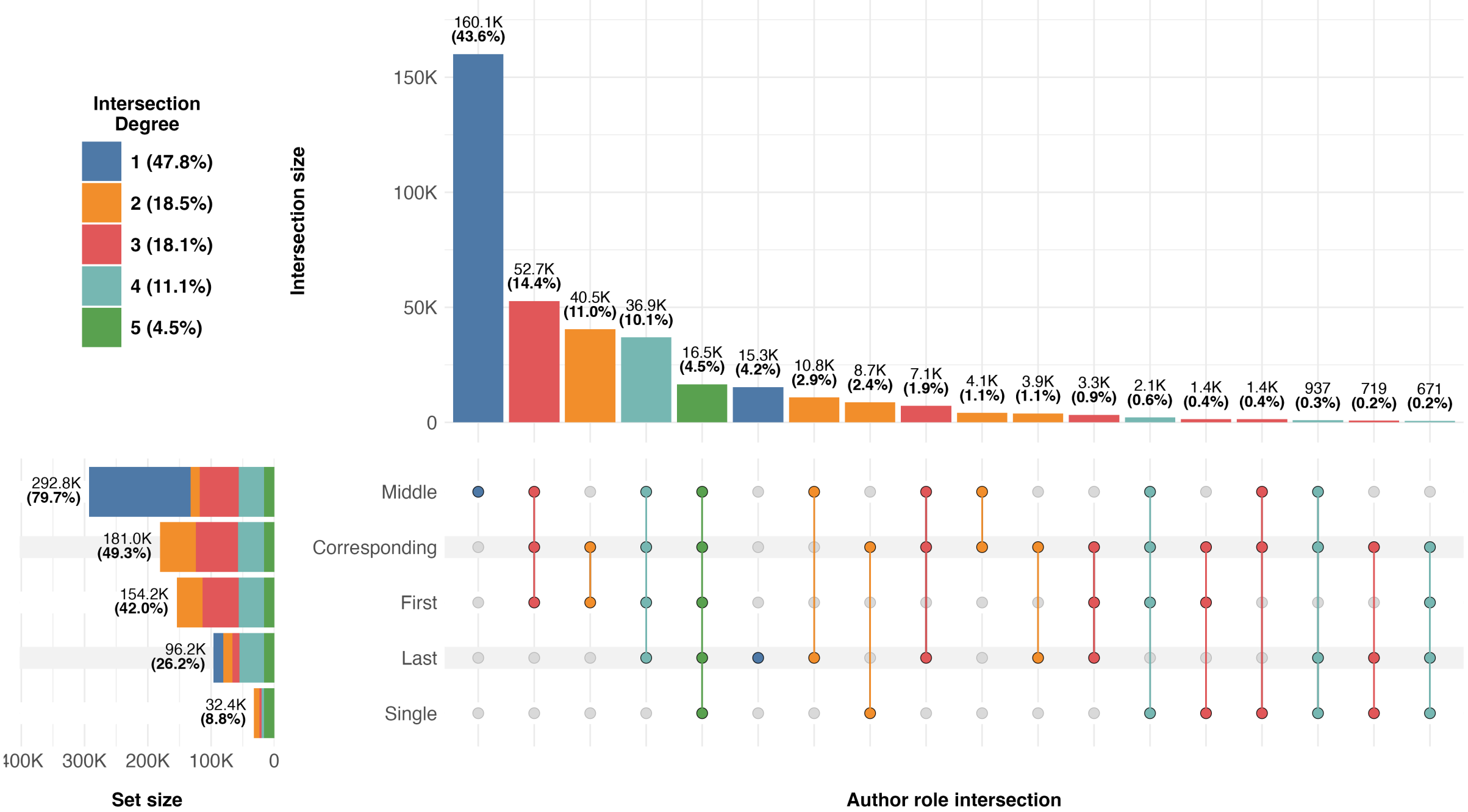

Figure 1. First name distribution across and between author role subsets. The horizontal bars of the set size subplot on the left give the total number of distinct first names in each author role subset. The matrix below the main panel encodes intersections: each column corresponds to a specific combination of subsets, with filled dots indicating which roles participate in that combination. The vertical bars of the intersection size subplot above the matrix give the number of first names belonging to each author role combination. Both plots as well as the matrix are color-coded by intersection degree, that is, by the number of distinct author roles included in each combination, ranging from first names unique to a single role to those shared across all five. Reading the columns thus reveals not only how many first names each role contributes, but also the extent to which first names are shared across roles or unique to a single one.

Given such high concentrations, however, one might suspect that they exceed what the uneven sizes of the author role subsets could produce on their own. To test this, we used a Monte Carlo permutation procedure for co-occurrence data (Kovács, 2014): in each of 10,000 iterations, every role was assigned a random set of first names equal in size to its observed subset, drawn independently of the other roles, thereby preserving the role marginals while removing any association between roles. Intersection proportions were computed over the names present in at least one role in each iteration, mirroring the structure of the observed data. Across all 10,000 iterations, every intersection degree fell outside its 99% interval under the random allocation model, and the departures followed a consistent pattern. Names confined to a single role (degree one) were sharply overrepresented at 47.8%, against an expected range of [25.3%, 25.6%], as were names shared across four roles (11.1% against [6.2%, 6.4%]) and all five (4.5% against [0.4%, 0.4%]); the intermediate degrees two and three were correspondingly depleted, at 18.5% and 18.1% against expected ranges of [41.4%, 41.8%] and [26.1%, 26.5%] respectively. The observed distribution is thus polarized relative to the random allocation model: first names concentrate at the two ends of the specialization scale, either tied to a single authorship position or shared across most or all of them, rather than spreading through the intermediate degrees as random allocation would predict. Authorship structure, not chance, thus shapes how names distribute across positions.

The subset sizes and intersection counts reported in Figure 1 reflect the composition of the WoS corpus after the three-step name-gender matching procedure rather than its raw onomastic structure. In particular, the decomposition of compound first names into their Wikidata components increases the effective number of distinct first names in each subset and may inflate intersection counts between roles, since the same compound name can be decomposed differently depending on which Wikidata components are found in each position. The figures in Figure 1 should therefore be read as reflecting onomastic diversity after gender assignment. This qualification does not, however, affect the random allocation model test: because that model preserves each role's observed subset size regardless of its origin, any inflation introduced by the decomposition affects the observed and expected distributions alike, and cannot account for the departure from the expected range. Whether this distributional structure also surfaces in how articles and citations spread across the genderedness spectrum will be the focus of the Analysis section.

### 3.2 Statistical Analysis

Following this data collection and matching procedure, the first name genderedness variable is constructed by dividing the male frequency of each first name by its total frequency. Through this procedure, all first names are assigned a continuous genderedness score between 0 and 1: a score of 1 indicates a name borne exclusively by male entities in Wikidata, a score near 0 a name borne almost exclusively by female entities, and intermediate values names shared across both. First names sharing the same score are then grouped together to form a single point along that interval, resulting in the creation of a graded genderedness spectrum containing 2,761 distinct values and forming an ordered axis $\mathrm{G}_1 < \mathrm{G}_2 < \ldots < \mathrm{G}_\mathrm{n}$ ranging from $4.528986 \times 10^{-4}$ to 1. By preserving this probability as a continuous score rather than collapsing it into a binary label, the measure aligns with Figlio's (2005, 2007) analysis of given names, the gender diagnosticity tradition in psychometrics (Lippa and Connelly, 1990; Lippa, 1991), and the continuous name-masculinity measure used in the Portia hypothesis literature (Coffey and McLaughlin, 2009).

Through this aggregation process, two analytically distinct, yet quantitatively connected, pairs of quantities are compiled at each point of the continuum. The first pair is lexical and draws on the Peircean distinction between tokens and types, here interpreted as the relation between the various occurrences of a first name and that first name itself, considered as an abstract shared form, without which different physical marks or sounds could not be recognized as being "the same sign" (Peirce, 1906/1933, CP 4.537). At each point of the continuum, types thus correspond to the number of distinct first names of a given genderedness score, while tokens represent the number of times those names occur. The second quantity pair is scientometric and turns from the names themselves to the scholarly activity attached to them. Within the context of the current project, an article is a publication credited to a name at a given point of the continuum, and a citation is a reference subsequently received by such a publication. At each point of the continuum, this pair opposes the number of articles credited to names of a given genderedness score, that is, their publication volume, to the number of citations those articles receive (their impact). Throughout, we use productivity as a shorthand for this name-level publication volume $\mathcal{A}$ at a given point of the spectrum, not in the career sense of an individual author's research output.

What connects this pair to the lexical one is the dual role the token plays across the two data sources. In Wikidata, a token is a human entity bearing a name, so that the token count reflects a name's demographic prevalence. In WoS, where the data record each first name of each author of each article, a token is an authorship occurrence which, from the standpoint of each name, is an article to its bearer's credit. The token is thus where the two pairs meet: counted against the diversity of names, it measures the lexical structure of the corpus; counted as scholarly output, it furnishes the article totals against which citations are weighed.

A brief example illustrates how both pairs are populated at a single point of the continuum. The first name 'Ranma', borne by one male and one female entity in Wikidata, receives a genderedness score of $1/(1 + 1) = 0.5$; 'Shinobu', borne by 66 male and 66 female entities, receives the same score ($66/132 = 0.5$) and is therefore grouped with 'Ranma' at that point, along with every other name for which male and female frequencies are equal. On the lexical side, these two names add 2 types and 134 tokens to the value 0.5 for the Wikidata database, which is associated with 603 different first names and 2,206 tokens in total; in the WoS dataset, 'Ranma' doesn't occur as an author first name, whereas Shinobu accounts for 467 of all 58,325 first name instances at that score. From a scientometric perspective, these 467 tokens correspond to the number of times the first name 'Shinobu' is found in the bylines of an article included in the dataset; because each such occurrence is an article credited to its bearer, they amount to 467 articles, which together drew 65,139 of the 37,534,322 citations received by names of genderedness 0.5.

As the numbers in this example and the subset sizes in Figure 1 show, the quantities considered in this project can span several orders of magnitude. Cumulative distributions offer a particularly well-suited analytical framework for handling dispersion and structure of this kind. Their adequacy first rests on the operationalization of genderedness itself, which yields a continuous, empirically grounded axis along which cumulative distributions can be built. The use of cumulative distributions is further motivated by the teachings of scientometrics itself. Decades of research have consistently shown that word, article, and citation frequencies distribute asymptotically across scholarly datasets according to specific power laws: Lotka's law for author productivity (Lotka, 1926), Bradford's law and cumulative advantage dynamics for citation concentration (Bradford, 1934; Price, 1976; Egghe, 2005), and Zipf's law for word frequencies, of which first names represent a special case (Zipf, 1949). Under such concentration, any fixed binning strategy would produce intervals of vastly unequal density, making comparisons across the genderedness spectrum arbitrary and sensitive to the choice of bin boundaries. Cumulative distributions bypass this problem entirely by representing the data without any artificial partitioning and allowing patterns of concentration to be read directly from the shape of the curve. Based on these considerations, four cumulative distributions are built along the genderedness axis, one for each of the quantities just introduced. At each value $G_i$, each gives the proportion of its corresponding total accounted for by first names whose genderedness does not exceed $G_i$: the cumulative share of types $\mathcal{T}$, of tokens $\mathcal{K}$, of articles $\mathcal{A}$, and of citations $\mathcal{C}$. All four are non-decreasing and reach 1 at $G_n$. To these is added a uniform reference $\mathcal{U}(G_i) = i/n$, which distributes weight equally across the n distinct values and will be of use in one of the scalar measures described below.

The comparative objectives of this study, however, call for a framework that describes not only how distributions cumulate along the genderedness spectrum, but also how they do so compared to others, revealing properties such as the relative shift in their centers of gravity, the amplitude of their divergence, and the asymmetry of each distribution along the spectrum. Such a framework is provided by the relative distribution methods of Handcock and Morris (1999), extended by Tygert (2021a, 2021b) and Kloumann et al. (2026) into a discrete formal framework based on cumulative difference functions. What makes the genderedness score well-suited as the ordering axis for this framework is precisely its probabilistic nature: Tygert (2021b) frames the independent variable of a cumulative difference plot as a score, of which a predicted probability is the canonical instance, while Handcock and Morris (1999) build their comparisons on the cumulative distribution of a reference variable. A probability that the bearer of a given first name is male is exactly such a score, ordering the entire analysis from maximally feminine to maximally masculine without any arbitrary discretization. Given this alignment, the cumulative difference function $\Delta_{\mathrm{i}}(\mathrm{A},\mathrm{B})$ is defined here as follows: for any two cumulative distributions $A$ and $B$ defined over the genderedness axis $G$, the cumulative difference function $\Delta_{\mathrm{i}}(\mathrm{A},\mathrm{B}) = \mathrm{A}(\mathrm{G_i}) - \mathrm{B}(\mathrm{G_i})$ traces, at each genderedness value $\mathrm{G_i}$, the extent to which $A$ has accumulated faster or slower than $B$ up to that point. By construction, the function begins and ends at zero: since both $A$ and $B$ reach 1 at $\mathrm{G_n}$, any surplus $A$ accumulates over $B$ in one region of the spectrum is necessarily offset elsewhere. The information lies not in the endpoints but in the shape of the curve between them, where a positive slope over an interval indicates that $A$ accumulates faster than $B$ across it, and a negative slope the reverse (Tygert, 2021b).

Three such cumulative difference functions, applied over the different author role types, structure the analyses that follow; in the sense formalized by Kloumann et al. (2026), each treats a given genderedness value as a matched pair of the two quantities compared. The first, $\Delta_{\mathrm{i}}(\mathcal{K},\mathcal{T})$, is lexical and compares the token distribution to the type distribution of both databases to each other. By measuring the degree to which first-name occurrences are concentrated on a small number of names relative to the overall diversity of names observed, this cumulative difference function characterizes the lexical structure of both datasets and provides the empirical basis for assessing the comparability of the Wikidata and WoS sources. The remaining two are bibliometric, specific to WoS, and broken down by disciplinary group. On the basis of the dual role of the token established earlier, $\Delta_i(\mathcal{A},\mathcal{T})$ carries the same analytical framework into a scientometric and disciplinary context, comparing scholarly output instead of first name occurrences against the type distribution. The third and last cumulative difference function, $\Delta_i(\mathcal{C},\mathcal{A})$, compares the cumulative distribution of citations to that of articles: by assessing impact regardless of productivity, it assesses the extent to which first name genderedness shapes citation distributions and as such represents the central analytical object of this project.

The cumulative difference functions are presented graphically in the analyses that follow, with curve shape as the primary source of information. For each function, three scalar measures complement the visual reading by making explicit distinct structural aspects of the curve: the location shift between the two distributions ($\lambda$), the amplitude of their divergence ($\nu$), and the polar concentration of each ($\eta$). The first, λ, locates the shift in center of gravity between the two distributions (Handcock and Morris, 1999). Letting $a(G_i)$ and $b(G_i)$ denote the local shares of distributions $A$ and $B$ at each genderedness value, that is, the proportion of each total located at $G_i$, $\lambda$ is the difference between their genderedness-weighted means: $\lambda = \sum i\, G_i \times a(G_i) - \sum i\, G_i \times b(G_i)$. As each data point of each distribution contributes to the mean proportionally to its share, $\lambda$ captures the average displacement of one distribution relative to the other along the spectrum: a positive $\lambda$ indicates that $A$ is, on average, shifted toward the masculine end relative to $B$, and a negative $\lambda$ toward the feminine end. To illustrate, consider two distributions assigning weights to the same ordered set of genderedness values but whose local shares follow respectively an arithmetic progression $(1/n, 2/n, 3/n, \ldots)$ and a geometric progression $(1/n, 2/n, 4/n, \ldots)$: once normalized to sum to 1, both cover the same range, but their weighted means differ substantially, as the geometric progression concentrates more mass toward the higher values, shifting its center of gravity upward relative to the arithmetic one. The second measure, $\nu = (\max_i \Delta_i(A,B) - \min \Delta_i(A,B))/2$, corresponds to the $V$ statistic of Kuiper (1962) projected into the $(0,1)$ interval. While the latter measures the total amplitude in difference between two cumulative distribution functions, $\nu$ returns half the difference between the maximum and minimum values of a cumulative difference function. In the current context, ν captures the extent of the relative polarization between the feminine and masculine extremes of the spectrum: the larger $\nu$ is, the wider the range between the peak and the trough of the cumulative difference function, and the more strongly the two distributions diverge at their respective extremes. The third measure, η, characterizes the polar concentration of a single distribution: the degree to which mass is concentrated at the masculine end relative to the uniform reference. Through its symbol, this scalar recalls the geometric interpretation of the h-index (Hirsch, 2005) obtained by means of the Lorenz curve, which it carries into a relative distributional setting (Egghe and Rousseau, 2019; Lorenz, 1905). It does so by locating the first point at which a distribution F crosses the anti-diagonal defined by the complement of the uniform cumulative distribution, $1 - \mathcal{U}(G_i)$: $\eta = 1 - \min\{G_i : F(G_i) \geq 1 - \mathcal{U}(G_i)\}$. Interpreted within the framework of the genderedness continuum, this measure marks the point beyond which the $\eta \times 100\%$ most masculinely-gendered names carry as much weight as all the remaining $(1 - \eta) \times 100\%$. While this measure has only recently been applied to scientometric contexts (Ghosh et al., 2014; Bérubé et al., 2018), the intuition it captures extends well beyond academic settings, as it underlies the familiar 80/20 rule (Juran, 1951; Pareto, 1896) and its most striking modern expression, the finding that the wealthiest 1% hold as much as the remaining 99% (Oxfam, 2016). Taken together, these three measures share a feature that underwrites their use throughout the analyses: each is bounded, $\lambda$ on $[-1, 1]$ and $\nu$ and $\eta$ on $[0, 1]$, so that none requires normalization and all can be compared directly across datasets, roles, and disciplinary groups.

# 4 Results

The procedures described in the previous section were designed to move the analysis from a binary inference of author sex to a continuous, cumulative, and relative measure of first name genderedness. This methodological shift was meant to resolve the two construct validity problems identified at the outset: the conflation of distinct individuals under a single inferred label, and the dichotomization of gender into two discrete categories. The first was addressed by aggregating both citation counts and gender frequencies at the level of the first name, which aligned the construct with its unit of measurement; the second, by preserving genderedness as a continuous score rather than a binary label, which represented gender as the graded quantity it is. This operationalization however carries an interpretive consequence: the analyses that follow bear on the masculinity and femininity of authors' first names, not on the maleness or femaleness of the authors themselves. This realignment in scope does not, however, deprive the results of all bearing on disparities between men and women: because genderedness is operationalized probabilistically as the probability that the bearer of a given first name is male, a statement about names of genderedness $G_i$ can be read as a statement about male authors with probability $G_i$ and about female authors with probability $1 - G_i$. In other words, at the extremes of the spectrum, where names are most strongly gendered and the probability that a name refers to an author of the opposite sex becomes vanishingly small, such correspondence becomes reliable enough to support inferences about sex-based differences.

### 4.1 First Names

Figure 2 shows the cumulative difference $\Delta_i(\mathcal{K}, \mathcal{T})$ between the token and type distributions along the genderedness spectrum, computed for both the Wikidata and WoS author role datasets. As can be seen from the outset, the five author role curves share a common shape:

each traces the same path along the genderedness axis, rising slightly at the feminine end of the continuum, then gradually descending toward the sharp rebound at its masculine extremity. As curves track the cumulative difference between tokens and types, this common trajectory across databases narratively translates into a consistent story across the spectrum: the slight rise at the feminine end marks a stretch where distinct names are few but each is borne often, popularity outrunning diversity; the long descent toward the masculine end marks the reverse, a proliferation of distinct names each comparatively rare; and the sharp rebound at the masculine extreme signals a return to concentration, where a handful of very common names dominate.

At first glance, this similarity of shape might seem to be an artifact of the method: after all, the lexicons of the Wikidata and WoS datasets underwent substantial transformation to be brought onto a shared genderedness axis, as first names were submitted to the same harmonization process and WoS first names were only kept if also present in the Wikidata lexicon. However, a shared axis does not in itself entail a shared shape; distributions defined over the same values may still accumulate along them very differently. More tellingly still and as Figure 1 showed, the five author-role subsets remain strikingly dissimilar despite these preprocessing steps: they differ markedly in size, in their degrees of pairwise overlap, and in their proportions of role-exclusive names, with intersection degrees polarized well beyond what chance allocation would produce. In our opinion, the fact that such heterogeneous subsets nevertheless yield congruent $\Delta_i(\mathcal{K},\mathcal{T})$ curves indicates that the lexical structure of genderedness, as embodied in the relationship between name diversity and name popularity along the spectrum, is a robust property of the scientific corpus as a whole, not an artifact of any similarity among the roles.

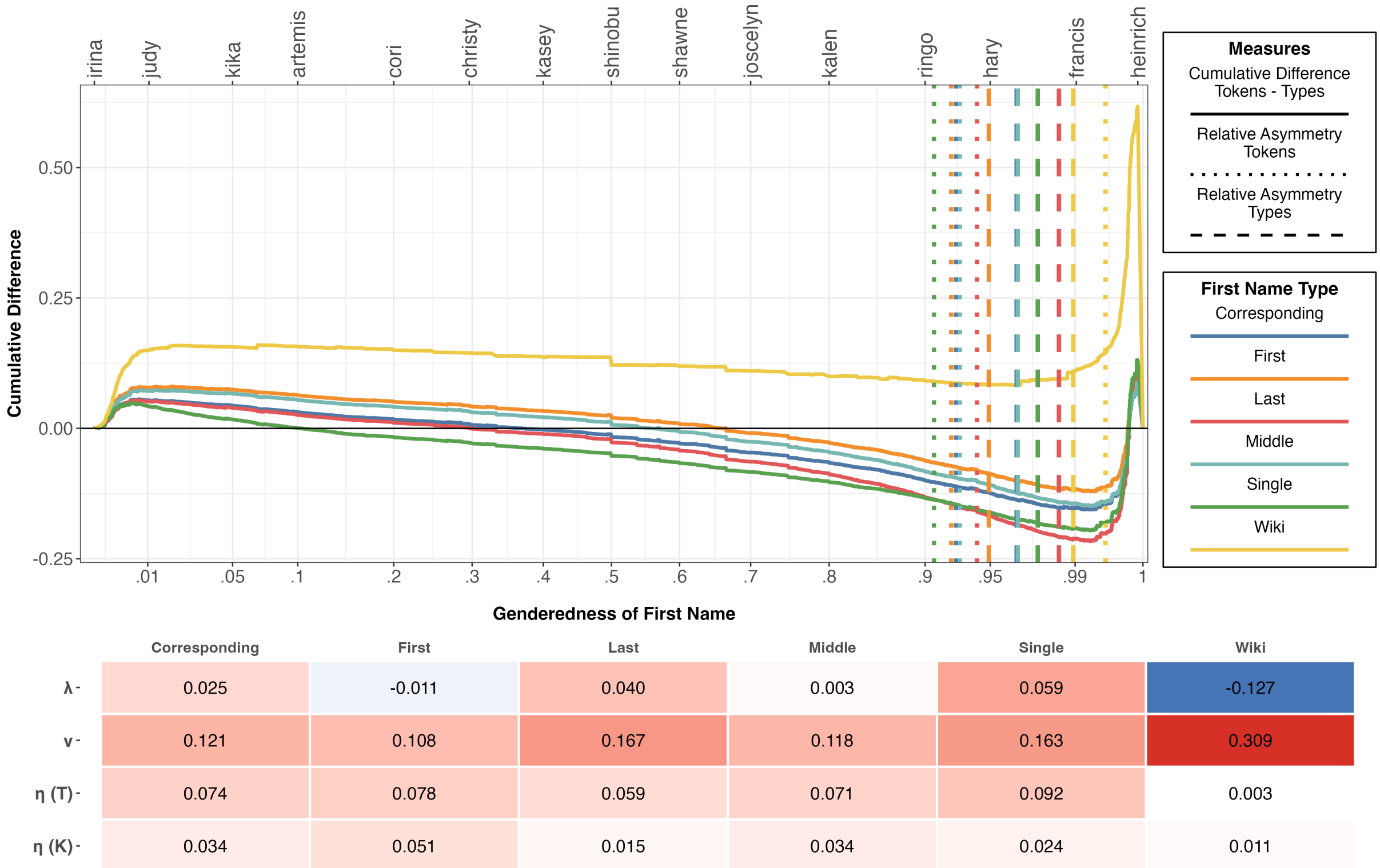

| | Corresponding | First | Last | Middle | Single | Wiki |
|---|---|---|---|---|---|---|
| λ | 0.025 | -0.011 | 0.040 | 0.003 | 0.059 | -0.127 |
| ν | 0.121 | 0.108 | 0.167 | 0.118 | 0.163 | 0.309 |
| η (T) | 0.074 | 0.078 | 0.059 | 0.071 | 0.092 | 0.003 |
| η (K) | 0.034 | 0.051 | 0.015 | 0.034 | 0.024 | 0.011 |

Figure 2. Cumulative Difference between first name types and tokens along the genderedness spectrum. Representative first names are placed at intervals along the x-axis as interpretive aids; the closer a name's genderedness score is to 0 or 1, the more femininely or masculinely gendered it is; in contrast, names near the midpoint (0.5) are not only the most gender-neutral or ambiguous, but also and often the least common ones, which makes first-name-based gender identification harder in that region of the spectrum. An arcsine transformation of the x-axis highlights distributional features at the lower and upper ends of the spectrum without distorting the ordinal relationship between data points. A scalar summary table beneath the figure reports $\eta(\mathcal{K}), \eta(\mathcal{T}), \nu$, and $\lambda$ for each of the six datasets (five WoS author roles and the Wikidata reference); $\lambda$ cells are shaded on a diverging blue–white–red scale (blue: feminine lean; red: masculine lean), while $\nu$, $\eta(\mathcal{T})$, and $\eta(\mathcal{K})$ follow a sequential white-to-red scale, with deeper shading indicating higher values.

Beyond this congruence of shape, the scalar measures offer a more nuanced portrait, one in which the WoS curves distinguish themselves structurally both in comparison with the Wikidata reference and amongst themselves. Against the Wikidata lexicon, the divergence is one of direction as much as degree. The lexicon's feminine lean ($\lambda = -0.127$) demarcates it from all WoS roles, a pattern consistent with sociological work documenting the more fashion-driven and historically variable nature of feminine first names (Lieberson, 2000). Its polar concentration also runs the other way: $\eta(\mathcal{T}) = 0.003$, an order of magnitude below any WoS role ($0.059 - 0.092$), meaning that a vanishingly small set of the most masculine first names accounts for as much diversity as all the rest. The Wikidata lexicon thus combines a feminine centre of gravity with a steep concentration at its masculine tip, a combination no role-specific distribution in the corpus displays; its amplitude ($\nu = 0.309$) is also the greatest of all, against a WoS range of $0.108 - 0.167$. Among the WoS curves themselves, the role distributions modulate the shared shape rather than depart from it. Amplitude ($\nu$) is the dimension along which they most clearly separate: last and single authors show the widest spread ($\nu = 0.167$ and $0.163$), while first and corresponding authors show the narrowest ($\nu = 0.108$ and $0.121$). Location ($\lambda$), by contrast, varies weakly and almost exclusively in degree rather than direction: corresponding, last, middle, and single authors all show a slight masculine lean ($\lambda = 0.025$, $0.040$, $0.003$, and $0.059$ respectively), with first authors the sole exception ($\lambda = -0.011$). The overall range remains modest at less than $0.07$ across all five roles, standing in sharp contrast with the Wikidata reference ($\lambda = -0.127$), which sits an order of magnitude further from parity.

Taken together, these findings establish the lexical architecture of the corpus: a structure that is remarkably stable across author roles, despite their substantial differences in size, overlap, and onomastic composition. This stability is not trivial: whatever asymmetries emerge in Figures 3 and 4 will reflect genuine differences in how articles and citations accumulate along the genderedness spectrum, not artifacts of role-specific lexical composition. One interpretive constraint nonetheless bears noting before turning to those analyses: the distributional readings rest on token counts that aggregate article occurrences across all authors bearing a given first name, without distinguishing the number of authors from their individual productivity. This precludes direct inference about individual-level productivity but not about scholarly output as such: because each token is itself an article, the concentration described here already reflects patterns of production, which the next section takes up explicitly by stratifying publication volume across disciplines through $\Delta_i(\mathcal{A}, \mathcal{T})$.

## 4.2 Articles

Figure 3 presents the cumulative difference $\Delta_i(\mathcal{A}, \mathcal{T})$ between the distribution of articles and first name types along the genderedness spectrum, stratified by author role and disciplinary group. Broken down by disciplinary group, the article distributions still trace the broad trajectory of Figure 2, a surge at the feminine extreme of the spectrum, a gradual descent through the middle range, and a final spike at the very masculine end. From that perspective, the disciplinary breakdown plays much the same role as the author role grouping: it introduces not rupture but nuance, a variation within a shape that the scalar measures shown under each subfigure articulate once more along the three dimensions of location ($\lambda$), amplitude ($\nu$), and polar concentration ($\eta$).

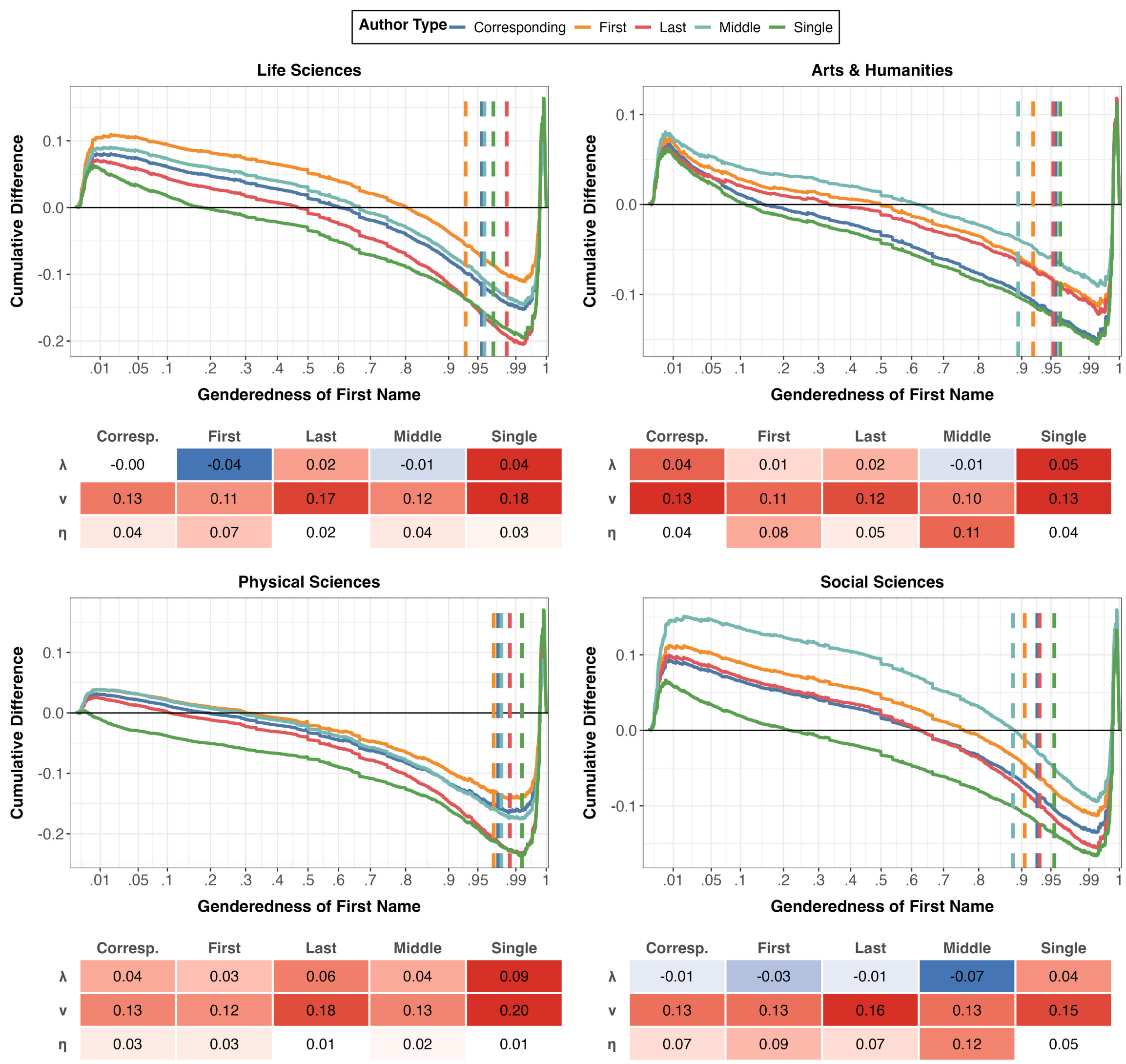

**Life Sciences**

| | Corresp. | First | Last | Middle | Single |
|---|---|---|---|---|---|
| λ | -0.00 | -0.04 | 0.02 | -0.01 | 0.04 |
| ν | 0.13 | 0.11 | 0.17 | 0.12 | 0.18 |
| η | 0.04 | 0.07 | 0.02 | 0.04 | 0.03 |

**Arts & Humanities**

| | Corresp. | First | Last | Middle | Single |
|---|---|---|---|---|---|
| λ | 0.04 | 0.01 | 0.02 | -0.01 | 0.05 |
| ν | 0.13 | 0.11 | 0.12 | 0.10 | 0.13 |
| η | 0.04 | 0.08 | 0.05 | 0.11 | 0.04 |

**Physical Sciences**

| | Corresp. | First | Last | Middle | Single |
|---|---|---|---|---|---|
| λ | 0.04 | 0.03 | 0.06 | 0.04 | 0.09 |
| ν | 0.13 | 0.12 | 0.18 | 0.13 | 0.20 |
| η | 0.03 | 0.03 | 0.01 | 0.02 | 0.01 |

**Social Sciences**

| | Corresp. | First | Last | Middle | Single |
|---|---|---|---|---|---|
| λ | -0.01 | -0.03 | -0.01 | -0.07 | 0.04 |
| ν | 0.13 | 0.13 | 0.16 | 0.13 | 0.15 |
| η | 0.07 | 0.09 | 0.07 | 0.12 | 0.05 |

Figure 3: Cumulative difference between articles and first name types, grouped by discipline. An arcsine transformation of the x-axis highlights distributional features at the lower and upper ends of the spectrum without distorting the ordinal relationship between data points. Taking the type distribution $\mathcal{T}$ as reference measures how scholarly output is distributed relative to the diversity of names available at each point of the spectrum: regions where $\Delta_i(\mathcal{A},\mathcal{T})$ is positive indicate that first names carrying that genderedness score publish disproportionately more articles than the diversity of names in that region would predict; regions where it is negative indicate the reverse. A scalar summary table beneath the figure reports $\eta(\mathcal{K})$, $\eta(\mathcal{T})$, $\nu$, and $\lambda$ for each of the six datasets (five WoS author roles and the Wikidata reference); $\lambda$ cells are shaded on a diverging blue–white–red scale (blue: feminine lean; red: masculine lean), while $\nu$, $\eta(\mathcal{T})$, and $\eta(\mathcal{K})$ follow a sequential white-to-red scale, with deeper shading indicating higher values.

Two of the three measures vary substantially across disciplinary groups, and in a disciplinarily ordered way. Location varies in direction as much as in degree: the physical sciences stand out with consistently positive $\lambda$ across all author roles ($0.028\ to\ 0.089$), a systematic masculine lean in article production that no other group displays across all roles, while the social sciences show the opposite tendency, with λ negative for four of five roles ($-0.075$ to $-0.007$), single authors excepted ($0.038$), indicating a consistent overrepresentation of femininely-gendered names in output. The life sciences and the arts and humanities occupy intermediate positions, with λ ranging from moderately negative to moderately positive in the former ($-0.040$ to $0.045$) and clustering near zero or slightly positive in the latter ($-0.006$ to $0.047$), neither showing a consistent directional tendency across roles. Polar concentration η orders the groups the same way: output is most sharply concentrated at the masculine tip in the physical sciences ($0.006$ to $0.030$) and least so in the social sciences ($0.045$ to $0.118$), the life sciences and arts and humanities falling in between. Amplitude is the exception. ν does take a range of values ($0.096$ to $0.204$ across all group-role combinations), but unlike the other two its variation carries no disciplinary signal: it never changes sign, it does not order the groups, and its extremes fall in unrelated cells, widest for single authors in the physical sciences and narrowest for middle authors in the arts and humanities. The departure of publication volume from name diversity is thus disciplinary in which end of the spectrum it favors ($\lambda$) and how steeply output concentrates there ($\eta$), but not in its sheer size ($\nu$).

These patterns carry a conceptually important implication: what disciplinary culture shapes is the direction and steepness of the departure from name diversity, not its magnitude. Which end of the spectrum benefits is disciplinary, the masculine end in the physical sciences and the feminine in the social sciences, and so is how sharply output concentrates there; the size of the departure is not. Analytically, this sets the baseline against which citation distributions will be evaluated: the question is no longer whether citations concentrate among masculinely-gendered names in absolute terms, but whether they concentrate there more than publication volume alone would predict.

### 4.3 Citations

Figure 4 presents the cumulative difference $\Delta_i(\mathcal{C}, \mathcal{A})$ between the distribution of citations and the distribution of articles along the genderedness spectrum, stratified by author role and disciplinary group. The most immediately striking feature here is that curves are predominantly negative across virtually the entire spectrum for all disciplinary groups and author roles, indicating a pervasive citation deficit for femininely- and neutrally-gendered names, consistent with prior work documenting gender disparities in citation practices (Larivière et al., 2013a; Maliniak et al., 2013; Caplar et al., 2017; Wu, 2023). At the masculine extreme, corresponding, first, and last authors show a notable upward recovery: a citation surplus for the most masculinely-gendered names that partially, but not fully, compensates for the deficit accumulated across the rest of the spectrum. The deficit at the feminine end is strongest for single authors, where the genderedness of an author's first name is not counterbalanced by that of any co-authors, and more attenuated for middle and last authors, whose first names attract comparatively less evaluative attention.

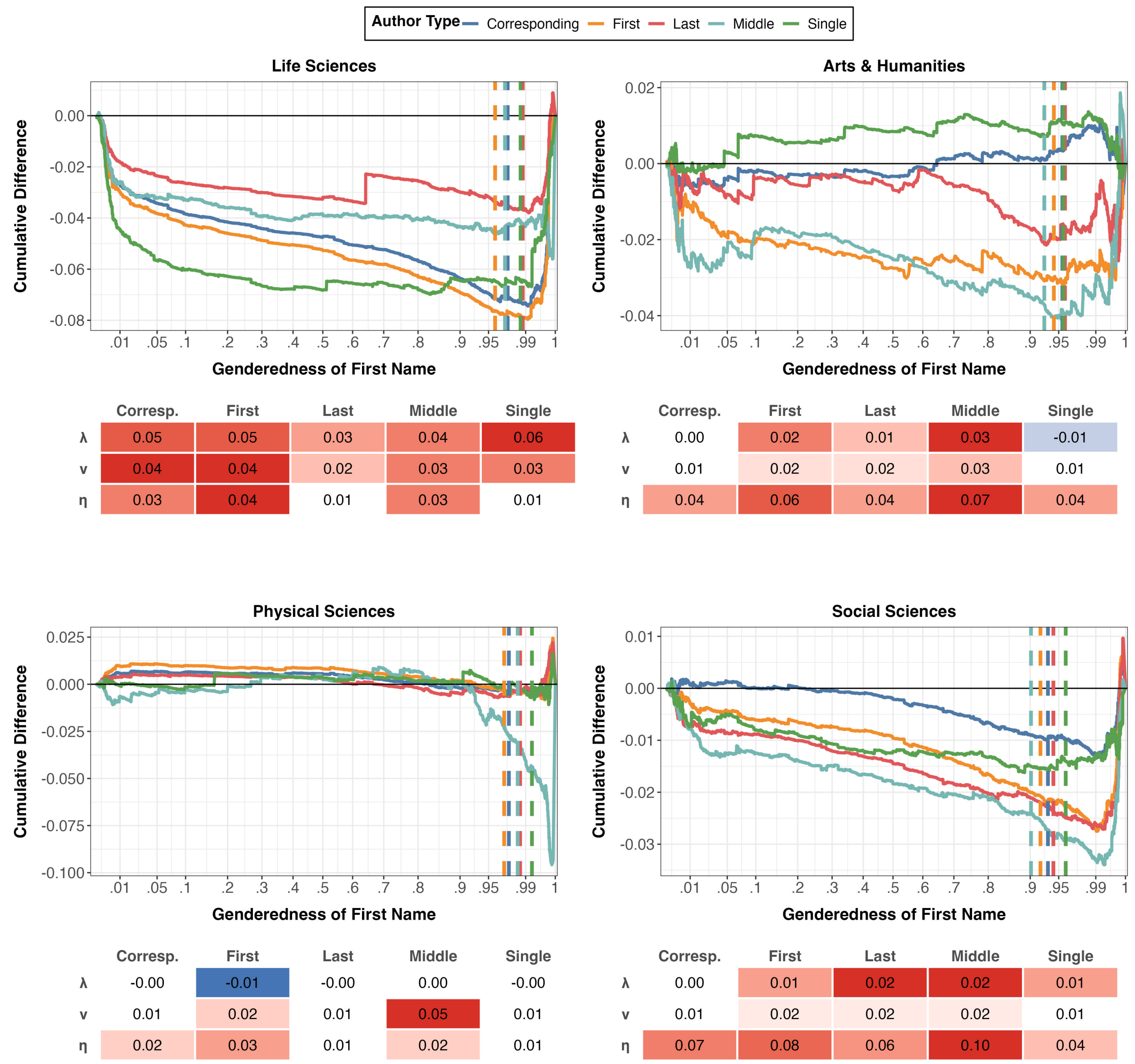

**Life Sciences**

| | Corresp. | First | Last | Middle | Single |
|---|---|---|---|---|---|
| λ | 0.05 | 0.05 | 0.03 | 0.04 | 0.06 |
| ν | 0.04 | 0.04 | 0.02 | 0.03 | 0.03 |
| η | 0.03 | 0.04 | 0.01 | 0.03 | 0.01 |

**Arts & Humanities**

| | Corresp. | First | Last | Middle | Single |
|---|---|---|---|---|---|
| λ | 0.00 | 0.02 | 0.01 | 0.03 | -0.01 |
| ν | 0.01 | 0.02 | 0.02 | 0.03 | 0.01 |
| η | 0.04 | 0.06 | 0.04 | 0.07 | 0.04 |

**Physical Sciences**

| | Corresp. | First | Last | Middle | Single |
|---|---|---|---|---|---|
| λ | -0.00 | -0.01 | -0.00 | 0.00 | -0.00 |
| ν | 0.01 | 0.02 | 0.01 | 0.05 | 0.01 |
| η | 0.02 | 0.03 | 0.01 | 0.02 | 0.01 |

**Social Sciences**

| | Corresp. | First | Last | Middle | Single |
|---|---|---|---|---|---|
| λ | 0.00 | 0.01 | 0.02 | 0.02 | 0.01 |
| ν | 0.01 | 0.02 | 0.02 | 0.02 | 0.01 |
| η | 0.07 | 0.08 | 0.06 | 0.10 | 0.04 |

*Figure 4: Cumulative difference between citations and articles, grouped by author type and disciplinary group. By subtracting the cumulative article distribution from the cumulative citation distribution, this analysis controls for the publication volume patterns documented above, isolating the degree to which citation counts are concentrated differently along the genderedness spectrum than articles themselves. A positive value of $\Delta_i(\mathcal{C}, \mathcal{A})$ at a given $G_i$ indicates a citation advantage: authors with names of that genderedness receive a disproportionate share of citations relative to their share of articles; inversely, a negative value indicates a citation deficit. As in Figure 3, scalar summary tables beneath each disciplinary plot report $\eta$, $\nu$, and $\lambda$ for the five author roles.*

The life sciences stand out as the disciplinary group with the strongest and most consistent citation asymmetry: $\lambda$ is positive for all author roles (ranging from 0.029 to 0.064), indicating a masculine citation advantage across the entire authorship structure, and $\nu$ ranges from 0.023 to 0.043, the highest values among non-middle author roles across all disciplinary groups. The physical sciences middle author role alone exceeds this range, with $\nu = 0.052$, for the structural reasons discussed above. The trough of the curve, which marks the point of maximum citation deficit for femininely-gendered names, falls at a genderedness of roughly 0.831 for single authors, well into the moderately masculine region of the spectrum and far from the extreme masculine end characterizing the productivity analysis.

By contrast, the physical sciences present a strikingly different pattern: $\lambda$ is near zero across all author roles (ranging from $-0.006$ to 0.000) and $\nu$ is consistently low (0.013 to 0.016), suggesting that the citation process in this disciplinary group is relatively neutral with respect to first name genderedness once publication volume is controlled for, a field-level contrast we take up in the Discussion. The notable exception is the middle author curve, which descends sharply toward the masculine end of the spectrum and whose amplitude ($\nu = 0.052$) constitutes by far the largest of any role-disciplinary group combination in the entire dataset. We interpret this as a structural artifact of mega-collaboration practices in high-energy physics, nuclear physics, and astrophysics. In such collaborations, individual articles may list hundreds or thousands of authors, the overwhelming majority of whom occupy middle-author positions (Wuchty et al., 2007; Milojević, 2010; Cronin, 2001). The masculine concentration of the middle-author peak likely reflects the gender composition of these large teams rather than evaluative bias in citing behavior, given the low representation of women in these subfields (Larivière et al., 2013a; Huang et al., 2020). This does not, however, dissolve the gender dimension of the pattern: the imbalance is simply located upstream, in the underrepresentation of women in these research communities, rather than in the citation process itself. This constitutes a structural hypothesis rather than a demonstrated causal mechanism, and the fact that first, last, and corresponding authors in the physical sciences show no such pattern is consistent with this reading.

Against the feminine lean documented in the productivity analysis, the social sciences show near-zero $\lambda$ across all citation roles (ranging from $-0.003$ to $0.018$), indicating that the overall citation distribution is roughly proportional to the article distribution along the genderedness spectrum. This apparent neutrality, however, conceals a sharp and deep trough: the maximum citation deficit concentrates at the extreme masculine end of the spectrum for most author roles, with only a modest upward recovery at the very tip. Single authors are the exception, with the trough falling well into the moderately masculine region rather than at the extreme. Amplitude is moderate across roles ($\nu$ from $0.018$ to $0.037$), comparable to the arts and humanities. The overall picture is one of a field where the feminine lean of scholarly output is not rewarded by a corresponding citation advantage, and where the most masculinely-gendered names retain a disproportionate share of citations that the productivity distribution alone does not account for.

Finally, Arts & Humanities fall between these two patterns in amplitude, but with a role-based polarization found nowhere else in the dataset. Single authors show a cumulative difference close to zero or slightly negative across virtually the entire spectrum ($\lambda = -0.008$). This reversal may reflect the different citation cultures of the humanities, where citation practices are less volume-driven, more selective, and less tied to large collaborative structures (Hicks, 2004; Larivière et al., 2006). First and last authors show a moderate masculine citation advantage ($\lambda = 0.024$ and $0.007$ respectively), and middle authors the strongest location shift in the disciplinary group ($\lambda = 0.026$), reflecting a sustained citation deficit for femininely-named authors in that role. The contrast between single authors, who show a marginal feminine advantage, and middle authors, who show the strongest masculine advantage, constitutes the sharpest role-based intra-disciplinary group polarization in the dataset.

Overall, the citation analyses reveal a pervasive asymmetry. Holding publication volume constant, that is, comparing each name's share of citations to its share of articles, femininely-gendered names accumulate citations at a deficit across virtually all disciplinary groups and author roles, while the most masculinely-gendered names collect a corresponding surplus. Yet this deficit is not confined to feminine and neutral names: it extends well into the masculine range, so that even moderately masculine names are under-cited, while the surplus is correspondingly narrow, confined to the most extremely masculine names alone. What the citation process rewards is therefore not masculinity as such but its unambiguous extreme.

This asymmetry is not uniform across disciplinary groups, though it varies in its structure rather than its amplitude, concentrating where citation volumes are highest and masculine names most concentrated. At the role level, the same ordering recurs wherever the shifts are large enough to read: the conventionally senior positions, corresponding and last author (Tscharntke et al., 2007), sit at the lower end of the range, while the strongest shift within each group falls to a non-senior role, single and first authors in the life sciences ($\lambda = 0.064$ and $0.054$) and middle authors in the social sciences ($\lambda = 0.018$, itself near zero). The gradient is thus carried chiefly by the life sciences, where the shifts are largest. This pattern suggests that established reputation attenuates the genderedness signal: senior authors typically enter citation decisions carrying an established track record, which reduces the salience of name genderedness as an evaluative cue, whereas first, middle, and single authors are more often unknown to the citing reader, leaving the genderedness signal with less to compete against. We offer this as an interpretation rather than a tested mechanism; it would be directly testable by conditioning on authors' prior citation stock or academic age, which the present name-level design does not capture.

The cumulative logic of the citation process plausibly does the rest. Because citation counts are themselves accumulated quantities, the snapshot already integrates whatever self-reinforcing dynamics have operated over each paper's life; what a cross-sectional design cannot show is the gap widening over time. The asymmetry documented here is therefore consistent with, rather than direct evidence of, a process in which small differences in recognition compound through cumulative advantage into sustained gaps.

# 5 Discussion

The present study yields two sets of findings, one methodological and one substantive. Methodologically, it shows that genderedness, operationalized as a continuous relative frequency-based measure of first name masculinity and femininity, represents a viable and analytically productive alternative to binary sex-based classifications in bibliometric research, one that resolves long-standing construct validity problems while remaining interpretable at the aggregate level. Substantively, this research documents a pervasive but structurally concentrated citation asymmetry: controlling for publication volume, femininely-gendered names accumulate citations at a deficit across virtually all disciplinary groups and author roles, while the most masculinely-gendered names collect a corresponding surplus. Crucially, this asymmetry is not uniform across the spectrum but concentrated at its masculine extreme: a broad deficit spanning the feminine and most of the masculine range, met by a surplus confined to its very tip, precisely where the genderedness signal is least ambiguous. Holding across disciplinary groups, even as its magnitude and direction vary substantially between them, this pattern invites interpretation through the mechanics of cumulative advantage that lie at the heart of scientometrics. Price (1976) showed that citation processes are self-reinforcing: already highly cited works attract further citations, so that small initial differences in recognition grow over time into large disparities. This dynamic, which Merton (1968) called the Matthew effect, has its gendered counterpart in the Matilda effect (Rossiter, 1993), whereby the same self-reinforcing process that advantages the already-recognized would disadvantage the already-overlooked along the axis of first name genderedness. The uneven concentration of citation recognition toward the masculine end is what such a process would produce; its specificity lies not in the fact of concentration but in what structures it.

Recent syntheses of the gender citation gap literature appear, at first glance, to tell a different story. In an empirical analysis of citation and salary data from Canadian universities, Wu (2023) shows that gender differences in career citations are largely mediated by research productivity and collaborative networks rather than by evaluative bias in citing behavior. In a subsequent synthesis of the gender citation gap literature, Wu (2024) then generalizes this conclusion across three distinct methodological approaches, finding that article-level gaps are consistently small or absent and that the robust disparity lies at the career level. The present findings appear to contradict this verdict: because $\Delta_i(\mathcal{C}, \mathcal{A})$ controls for publication volume and operates on citations relative to articles, it is closest in spirit to the per-article approach, yet it documents a pervasive feminine deficit where that approach typically finds none. The apparent contradiction, however, reflects a difference in design rather than a genuine empirical divergence. The per-article studies synthesized by Wu compare two binary author categories, collapsing the entire genderedness spectrum into a single contrast. A polarized asymmetry is precisely what such a design is least equipped to detect: by pooling names at the extreme masculine end together with all moderately masculine names, it averages a strong effect with a weak or absent one, attenuating the divergence concentrated at the tips. The apparent absence of an article-level gap in the binary literature is thus, on this reading, partly an artifact of dichotomization itself, the very construct-validity problem the present operationalization was designed to address.

In our view, both this problematic operationalization and the citation asymmetries documented here may share a common cognitive origin: the principles underlying categorical perception. When a category boundary is learned, perception of the underlying continuum is not merely labeled but actively deformed: within-category distances are compressed and between-category distances are amplified on either side of the boundary, such that members of the same category come to seem more alike and members of different categories more dissimilar than the underlying signal warrants (Harnad, 1987; Livingston, Andrews and Harnad, 1998). What makes this framework particularly apt here is that it grounds categorization in behavior rather than in conscious representation: just as a bee's differential response to flowers requires no explicit concept of flower, a citing author's differential response to strongly masculine names requires no explicit reasoning about gender, only the automatic activation of a behavioral regularity stabilized by prior experience with the category (Rosch, 1975, 1978). This same cognitive structure, we argue, operates at two levels in the present context. At the level of the citing author, the operative categorical structure is not a symmetric male/female binary but a prototypical asymmetry. That the prototype of the scientist or scholar is strongly masculine is well documented: children asked to draw a scientist predominantly draw a man (Miller et al., 2018), the lay concept of scientist overlaps far more with stereotypes of men than of women (Carli et al., 2016), and science is implicitly associated with maleness across cultures (Nosek et al., 2009). Our conjecture is that this prototype is activated through first name genderedness, with male genderedness functioning as the unmarked default (Bem, 1993) against which feminine, androgynous, and ambiguous names register as marked departures. The operative distinction is thus not male versus female: only names that unambiguously activate the masculine prototype receive the full citation premium it confers; everything else is residual. Under the low-deliberation conditions that characterize citation behavior, this perceptual asymmetry operates below the threshold of conscious deliberation (Greenwald and Banaji, 1995), such that the citer need not explicitly reason about gender for the categorical structure to shape the evaluation. The distribution of citation recognition thus maps onto the distribution of prototypical salience: strongest where the masculine prototype is most clearly activated, attenuated where it dissolves into ambiguity. At the level of the bibliometric researcher, the same categorical structure may have operated differently but no less consequentially: not as an active distortion of perception, but as an invisible default. Like the native speaker who does not perceive the grammar through which they produce language, researchers who operationalize gender as a binary may not recognize their instrument as a theoretical choice; the binary functions less as a conviction than as a blind spot, reproduced without ever registering as a position one could hold or reject (Bem, 1993; Keyes, 2018). The binary choice was invisible precisely because categorical perception had made it seem natural. The present framework's departure from that operationalization is, in this light, a deliberate resistance to the very cognitive reflex it sets out to study.

This account also clarifies how individual tendencies of such modest magnitude can produce disparities of the scale documented here, and it does so at both levels identified above. Schelling (1978) showed that the relation between individual motives and collective outcomes is rarely one of resemblance: mild and uncoordinated individual preferences, each far too weak to count as a conviction, can aggregate into pronounced and durable macro-level patterns that no one intended. At the level of the citing author, the present case carries this logic below the level of preference itself, since the perceptual asymmetry posited here need not rise to a motive at all. If mild preferences already suffice to produce sharp aggregate divisions, a sub-deliberative bias diffused across the citing population can do the same, even more so once the cumulative dynamics of citation convert each small and repeated increment of recognition into a widening gap. At the level of the bibliometric researcher, the same disproportion governs a different outcome: no one set out to render the genderedness gradient invisible, yet the binary instrument became the field's default through the accumulation of choices each unremarkable to the one making it, until the convention itself foreclosed the very pattern it could not register. In neither case, however, does the relation between soft, innocuous micromotives and stark, unintended macro-outcomes run in reverse, whether the harm is distributive, as in the citation deficit, or epistemic, as in the erasure of a gradient the field could not see. The aggregate citation asymmetry licenses no inference about the intent, or even the awareness, of any individual citer (Traag and Waltman, 2022), just as the dominance of the binary instrument licenses none about the intent, or even the awareness, of any individual researcher. In both, the macro-level outcome is real and consequential without implying that anyone chose it.

This account is offered as interpretation rather than demonstration: it unifies the construct-validity concerns raised at the outset with the empirical patterns documented here and could be tested experimentally. On that note, convergent evidence already exists in the citation record itself. Analyzing more than one million articles published across thirty-five physics journals over twenty-five years, Teich et al. (2022) construct a citation model that predicts each paper's expected citation count from a set of characteristics independent of author gender, then compare observed to expected citations as a function of authorship. They find that papers authored by women are systematically under-cited relative to expectation, while papers authored by men are systematically over-cited, and, crucially, that this imbalance is strongest in two conditions: when the citing paper appears in a general physics venue rather than a specialized subfield journal, and when the citing authors are unlikely to be familiar with the cited work or its authors. These are precisely the low-familiarity conditions under which a name-based heuristic would be expected to exert its greatest influence: when the citing reader cannot draw on direct knowledge of the cited work's content, authors, or community, the first name becomes a more salient signal, and the implicit-bias mechanism posited here operates with the fewest competing inputs.

This conditional and structural character also reconciles the mechanism with the disciplinary heterogeneity documented above. Because $\Delta_i(\mathcal{C}, \mathcal{A})$ isolates the amplification a pattern receives at the citation stage, over and above its expression in production, a field whose masculine concentration is already carried by its article distribution will show little residual at the citation stage even if the underlying disposition is no weaker there. The near-zero physical-sciences residual is, on this reading, not the absence of gender asymmetry but its location upstream, in who publishes rather than in who is cited, the same displacement invoked above for the middle-author peak. A decision-level disposition constant across fields is thus fully compatible with a distributional footprint that is not, since whether that disposition leaves a measurable residual depends on how much of the concentration production has already absorbed, on citation volume, and on the genderedness variance available for it to act on. The low-familiarity condition identified by Teich et al. (2022) would further modulate this within each field, and plausibly between fields to the extent their collaborative structures differ, though disentangling a disposition that is genuinely weaker from one whose footprint this design renders invisible would require the familiarity and network data the present name-level approach does not capture.

Several constraints bear on these conclusions. The genderedness scores computed here reflect the composition of Wikidata's predominantly Western and English-language biographical entries (Konieczny and Klein, 2018; Tripodi, 2023), and names from non-Western naming traditions may carry scores that poorly represent local conventions. This concern is partly attenuated by the comparable geographic and linguistic skew of the WoS Core Collection itself (Mongeon and Paul-Hus, 2016; Van Leeuwen et al., 2001; Larivière et al., 2013a), which reduces the mismatch between the two sources without eliminating it. Another related concern however, one which the present design does not address, is that part of the asymmetry attributed here to genderedness may in fact reflect correlated perceptions of name foreignness. Author self-citations are also included and, in a name-level design, cannot be separated from citations between distinct authors who share a name. Because self-citation is more frequent among men (Azoulay and Lynn, 2020), masculinely-gendered names plausibly receive a larger share of self-citations, inflating their citations relative to their articles at the masculine end and contributing to the surplus observed there. Self-citations are, however, a minority of all citations, and the surplus is confined to the extreme masculine tip, so this mechanism can contribute to that surplus without fully accounting for it. Finally, because strongly gendered names are unevenly distributed across birth cohorts, part of the asymmetry at the masculine extreme may reflect the gendered career-progression dynamics documented at the author level (Wu, 2023), whereby seniority, accumulated reputation, and longer publication histories rather than the perception of the name itself drive citation accumulation.

These limitations point toward several directions for future research. First and foremost, the causes of the observed asymmetries remain to be determined. Reader perception, network positioning, and disciplinary composition each represent plausible drivers, and the distributional framework developed here provides a natural scaffold for distinguishing between them, as each would leave a distinct signature across the genderedness spectrum. On a related note, the perceptual account developed here could be tested directly through experimental work targeting how citing researchers perceive and respond to author names of varying genderedness. Perhaps more importantly, replicating this analysis in national scholarly communities outside the US, with name-gender frequency data drawn from culturally appropriate sources, would test whether the polarization documented here is a property of citation behavior in general or one shaped by the specific context analyzed here. Empirical support for this extension already exists: across 5,500 communication scholars in eleven countries, Rajkó et al. (2023) document that the magnitude of the citation gap varies substantially with national context, consistent with the expectation that name-based citation effects track local onomastic conventions. Pursuing this lead would require name-gender frequency data calibrated to each national context, since name-to-gender inference services perform substantially less well for non-Western names (Santamaría and Mihaljević, 2018). Wikidata's entity-level structure offers one possible affordance here, as genderedness scores could in principle be computed within specific national contexts by aggregating name-sex frequencies by country of birth but fully exploiting it would require knowing the cultural background of both cited and citing authors, a methodological requirement that would substantially transform the study design.

## 5.1 Open science practices

All gender-related data used in this research has been extracted from semantic triples contained in the Wikidata knowledge base, whose "truthy" and complete data dumps are freely available at https://dumps.wikimedia.org/wikidatawiki/entities/. The data processing pipeline, including all Python scripts for data extraction, name matching, and scalar computation, is available at https://github.com/msaintemarie/tom_dick_harry.git. The processed dataset is archived on Zenodo at https://zenodo.org/records/21029716.

## 5.2 Author contributions

Conceptualization: MHSM
Data curation: MHSM, VL
Formal analysis: MHSM
Funding acquisition: VL
Investigation: MHSM
Methodology: MHSM
Project administration: VL
Software: MHSM
Resources: VL
Supervision: VL
Validation: MHSM, VL
Visualization: MHSM
Writing – original draft: MHSM
Writing – review & editing: MHSM, VL

## 5.3 Funding information

This work was supported by the Social Science and Humanities Research Council of Canada Pan-Canadian Knowledge Access Initiative Grant (Grant 1007-2023-0001), and the *Fonds de recherche du Québec - Société et Culture* through the *Programme d'appui aux Chaires UNESCO* (Grant 338828).

## 5.4 References

Amering, M., Schrank, B., & Sibitz, I. (2011). The gender gap in high-impact psychiatry journals. *Academic Medicine, 86*(8), 946–952.

Ansara, Y. G., & Hegarty, P. (2014). Methodologies of misgendering: Recommendations for reducing cisgenderism in psychological research. *Feminism & Psychology, 24*(2), 259–270.

Aura, S., & Hess, G. D. (2004). *What's in a name* (Working Paper No. 1190). Center for Economic Studies; ifo Institute (CESifo).

Aura, S., & Hess, G. D. (2010). What's in a name? *Economic Inquiry, 48*(1), 214–227.

Azoulay, P., & Lynn, F. B. (2020). Self-citation, cumulative advantage, and gender inequality in science. Sociological Science, 7, 152–186. https://sociologicalscience.com/articles-v7-7-152.

Bauer, P. J., & Coyne, M. J. (1997). When the name says it all: Preschoolers' recognition and use of the gendered nature of common proper names. *Social Development, 6*(3), 271–291.

Bem, S. L. (1987). Masculinity and femininity exist only in the mind of the perceiver. In J. M. Reinisch, L. A. Rosenblum, & S. A. Sander (Eds.), *Masculinity/Femininity: Basic perspectives* (pp. 304–311). Oxford University Press.

Bendels, M. H. K., Müller, R., Brueggmann, D., & Groneberg, D. A. (2018). Gender disparities in high-quality research revealed by Nature Index journals. *PLOS ONE, 13*(1), e0189136. https://doi.org/10.1371/journal.pone.0189136

Bengtsson, R., Iverman, E. & Hinnerich, B.T. (2012). Gender and Ethnic Discrimination in the Rental Housing Market. *Applied Economics Letters*, 19 (1): 1–5.

Bérubé, N., Ghiasi, G., Sainte-Marie, M., & Larivière, V. (2020). Wiki-Gendersort: Automatic gender detection using first names in Wikipedia.

Bérubé, N., Sainte-Marie, M., Mongeon, P., & Larivière, V. (2018). Words by the tail: Assessing lexical diversity in scholarly titles using frequency-rank distribution tail fits. *PloS ONE*, *13*(7), e0197775. https://doi.org/10.1371/journal.pone.0197775

Biederman, I., & Shiffrar, M. M. (1987). Sexing day-old chicks: A case study and expert systems analysis of a difficult perceptual-learning task. *Journal of Experimental Psychology: Learning, Memory, and Cognition*, 13(4), 640–645. (doi:10.1037/0278-7393.13.4.640)

Blevins, C., & Mullen, L. (2015). Jane, John ... Leslie? A historical method for algorithmic gender prediction. *Digital Humanities Quarterly, 9*(3).

Boekhout, H., van der Weijden, I., & Waltman, L. (2021). Gender differences in scientific careers: A large-scale bibliometric analysis. arXiv:2106.12624.

Bornmann, L., and Daniel, H.-D. (2008). What Do Citation Counts Measure? A Review of Studies on Citing Behavior. *Journal of Documentation*, 64(1): 45–80. https://doi.org/10.1108/00220410810844150.

Bornstein, K. (1994). *Gender outlaw: On Men, Women, and the Rest of Us*. New York: Vintage Books.

Bradford, S. C. (1934). Sources of information on specific subjects. *Engineering*, 137, 85–86.

Bright, W. (2003). What is a name? Reflections on onomastics. *Language and Linguistics, 4*(4), 669–681.

Budden, A. E., Tregenza, T., Aarssen, L. W., Koricheva, J., Leimu, R., & Lortie, C. J. (2008). Double-blind review favours increased representation of female authors. *Trends in Ecology & Evolution, 23*(1), 4–6.

Butler, J. (1990). Gender Trouble. Routledge.

Caplar, N., Tacchella, S., & Birrer, S. (2017). Quantitative evaluation of gender bias in astronomical publications from citation counts. *Nature Astronomy, 1*, 0141. https://doi.org/10.1038/s41550-017-0141.

Carli, L. L., Alawa, L., Lee, Y., Zhao, B., & Kim, E. (2016). Stereotypes about gender and science: Women≠ scientists. *Psychology of Women Quarterly*, *40*(2), 244-260.

Coffey, B., & McLaughlin, P. A. (2009). Do masculine names help female lawyers become judges? Evidence from South Carolina. *American Law and Economics Review, 11*(1), 112–133.

Coffey, B., & McLaughlin, P. A. (2016). The effect on lawyers' income of gender information contained in first names. *Review of Law & Economics, 12*(1), 57–76.

Costas R., Nane T. & Lariviere V. (2015), Is the Year of First Publication a Good Proxy of Scholars' Academic Age? Proceedings of the 15th International Society of Scientometrics and Infometrics Conference, Istanbul, Turkey, 29 June 2015 - 4 July 2015: 988-998.

Cronin, B. (2001). Hyperauthorship: A postmodern perversion or evidence of a structural shift in scholarly communication practices? *Journal of the American Society for Information Science and Technology, 52*(7), 558–569. https://doi.org/10.1002/asi.1097

Del Giudice M. 2024. Statistical indices of masculinity-femininity: A theoretical and practical framework. Behavior Research Methods 56: 6538–6556.

Egghe, L. (2005). Power Laws in the Information Production Process: Lotkaian Informetrics. Elsevier.

Egghe, L., & Rousseau, R. (2019). A geometric relation between the h-index and the Lorenz curve. *Scientometrics*, 119(2), 1281–1284. https://doi.org/10.1007/s11192-019-03083-2

Engqvist, L., & Frommen, J. G. (2008). Double-blind peer review and gender publication bias. *Animal Behaviour*, *76*(3).

Fei, Q., Bell, R. L., Kennebrew, D., & Minton, S. (2023). The Career Benefit of Having a "Good" Name. Journal of Applied Business & Economics, 25(4).

Figlio, D. N. (2005). Why Barbie says "math is hard." National Bureau of Economic Research. https://users.nber.org/~confer/2005/hiedf05/figlio.pdf

Figlio, D. N. (2007). Boys named Sue: Disruptive children and their peers. *Education Finance and Policy, 2*(4), 376–394.

Foschi, M., and Valenzuela, J. (2012). Who Is the Better Applicant? Effects from Gender, Academic Record, and Type of Decision. *Social Science Research*, 41(4): 949–64.

Fox, M.F. (2001). Women, science, and academia: Graduate education and careers. *Gender & Society*, 15(5), 654-666.

Ghosh, A., Chattopadhyay, N., & Chakrabarti, B. K. (2014). Inequality in societies, academic institutions and science journals: Gini and k-indices. *Physica A*, 410, 30–34.

Gilbert, N. (1977). Referencing as Persuasion. *Social Studies of Science*, 7(1): 113–122. https://doi.org/10.1177/030631277700700112

Greenwald, A.G., and Banaji, M.R. (1995). Implicit Social Cognition: Attitudes, Self-Esteem, and Stereotypes. *Psychological Review*, 102(1): 4–27.

Goldin, C., & Rouse, C. (2000). Orchestrating impartiality: The impact of "blind" auditions on female musicians. *American Economic Review, 90*(4), 715–741.

González-Salmón, E., Chinchilla-Rodríguez, Z., & Robinson-Garcia, N. (2024). *The woman's researcher tale: A review of bibliometric methods and results for studying gender in science* (U-CHASS White Papers No. 1). https://doi.org/10.5281/zenodo.10590300

Halevi, G. (2019). Bibliometric studies on gender disparities in science. In W. Glänzel, H. F. Moed, U. Schmoch, & M. Thelwall (Eds.), *Springer handbook of science and technology indicators* (pp. 563–580). Springer.

Handcock, M. S., & Morris, M. (1999). Relative Distribution Methods in the Social Sciences. Springer.

Hamidi, F., Scheuerman, M. K., and Branham, S. M. (2018). "Gender recognition or gender reductionism?: the social implications of embedded gender recognition systems," in Proceedings of the 2018 CHI Conference on Human Factors in Computing Systems, CHI '18 (New York, NY: ACM), 8:1–8:13.

Harnad, S. (1987). Psychophysical and cognitive aspects of categorical perception: A critical overview. In S. Harnad (Ed.), Categorical Perception: The Groundwork of Cognition (pp. 1–52). Cambridge University Press.

Hicks, D. M. (2004). The four literatures of social science. In H. F. Moed, W. Glänzel, & U. Schmoch (Eds.), *Handbook of quantitative science and technology research* (pp. 473–496). Springer. https://doi.org/10.1007/1-4020-2755-9_22

Hirsch, J. E. (2005). An index to quantify an individual's scientific research output. *Proceedings of the National Academy of Sciences*, 102(46), 16569–16572.

Hough, C. (2000). Towards an Explanation of Phonetic Differentiation in Masculine and Feminine Personal Names. *Journal of Linguistics*, 36 (1): 1–11.

Juran, J. M. (1951). Quality Control Handbook. New York: McGraw-Hill.

Keyes, O. (2018). The misgendering machines: Trans/HCI implications of automatic gender recognition. *Proceedings of the ACM on human-computer interaction*, *2*(CSCW), 1-22.
Keller, S. E. (1999). Operations of Legal Rhetoric: Examining Transsexual and Judicial Identity. *Harvard Civil Rights-Civil Liberties Law Review (CR-CL)*, *34*(2), 329-384.
Kloumann, I., Korevaar, H., McConnell, C., Tygert, M., & Zhao, J. (2026). Cumulative differences between paired samples. *Statistical Analysis and Data Mining: An ASA Data Science Journal*, *19*(1), e70067.
Kong, H., Martin-Gutierrez, S., & Karimi, F. (2022). Influence of the first-mover advantage on the gender disparities in physics citations. *Communications Physics*, 5(1), Article 1. https://doi.org/10.1038/s42005-022-00995-z
Konieczny, P., & Klein, M. (2018). Gender gap through time and space: A journey through Wikipedia biographies via the Wikidata human gender indicator. *New Media & Society, 20*(12), 4608–4633.
Kovács, B. (2014). A Monte Carlo permutation test for co-occurrence data. *Quality & Quantity*, 48(2), 955–960. https://doi.org/10.1007/s11135-012-9817-x
Krishan, K., Chatterjee, P. M., Kanchan, T., Kaur, S., Baryah, N., & Singh, R. K. (2016). A review of sex estimation techniques during examination of skeletal remains in forensic anthropology casework. Forensic Science International, 261, 165.e1–165.e8. doi:10.1016/j.forsciint.2016.02.007 (PMID: 26926105).
Kuiper, N. H. (1960). Tests concerning random points on a circle. Proceedings of the Koninklijke Nederlandse Akademie van Wetenschappen, Series A, 63, 38-47.
Lammers, J., Gordijn, E.H., and Otten, S. (2009). Iron Ladies, Men of Steel: The Effects of Gender Stereotyping on the Perception of Male and Female Candidates Are Moderated by Prototypicality. *European Journal of Social Psychology*, 39(2): 186–95.
Larivière, V., Archambault, É., Gingras, Y., & Vignola-Gagné, É. (2006). The place of serials in referencing practices: Comparing natural sciences and engineering with social sciences and humanities. *Journal of the American Society for Information Science and Technology, 57*(8), 997–1004. https://doi.org/10.1002/asi.20349
Larivière, V., Vignola-Gagné, E., Villeneuve, C., Gélinas, P., & Gingras, Y. (2011). Sex differences in research funding, productivity and impact: an analysis of Québec university professors. *Scientometrics*, *87*(3), 483-498.
Larivière, V., Desrochers, N., Macaluso, B., Mongeon, P., Paul-Hus, A., & Sugimoto, C. R. (2016). Contributorship and division of labor in knowledge production. *Social Studies of Science, 46*(3), 417–435. https://doi.org/10.1177/0306312716650046
Larivière, V., Ni, C., Gingras, Y., Cronin, B., & Sugimoto, C. R. (2013a). Bibliometrics: Global gender disparities in science. *Nature, 504*(7479), 211–213.
Larivière, V., Ni, C., Gingras, Y., Cronin, B., & Sugimoto, C. R. (2013b). Supplementary information to: Global gender disparities in science. *Nature, 504*(7479).
Leahey, E. (2006). Gender differences in productivity: Research specialization as a missing link. *Gender & Society*, *20*(6), 754-780.
Lerchenmueller, M. J., & Sorenson, O. (2018). The gender gap in early career transitions in the life sciences. *Research Policy, 47*(6), 1007–1017. https://doi.org/10.1016/j.respol.2018.02.009.
Lex, A., Gehlenborg, N., Strobelt, H., Vuillemot, R., & Pfister, H. (2014). UpSet: visualization of intersecting sets. *IEEE transactions on visualization and computer graphics*, *20*(12), 1983-1992.
Lieberson, S. (2000). *A matter of taste: How names, fashions, and culture change.* New Haven, CT: Yale University Press.
Lindqvist, A., Gustafsson Sendén, M., & Renström, E. A. (2021). What is gender, anyway: A review of the options for operationalising gender. *Psychology & Sexuality, 12*(4), 332–344.

Lippa, R. (1991). Some psychometric characteristics of gender diagnosticity measures: Reliability, validity, consistency across domains, and relationship to the Big Five. Journal of Personality and Social Psychology, 61(6), 1000.

Lippa, R., & Connelly, S. (1990). Gender diagnosticity: A new Bayesian approach to gender-related individual differences. Journal of Personality and Social Psychology, 59(5), 1051.

Litwiller, F. (2020). Normative drag culture and the making of precarity. *Leisure Studies*, 39, 4, 600–612.

Livingston, K. R., Andrews, J. K., & Harnad, S. (1998). Categorical perception effects induced by category learning. Journal of Experimental Psychology: Learning, Memory, and Cognition, 24(3), 732–753.

Lockhart, J. W., King, M. M., & Munsch, C. (2023). Name-based demographic inference and the unequal distribution of misrecognition. *Nature Human Behaviour*, *7*(7), 1084-1095.

Lorenz, M. O. (1905). Methods of measuring the concentration of wealth. *Publications of the American Statistical Association*, *9*(70), 209–219. https://doi.org/10.2307/2276207

Lotka, A. J. (1926). The frequency distribution of scientific productivity. *Journal of the Washington Academy of Sciences*, 16(12), 317–323.

Mackay, F. (2021). Female Masculinities and the Gender Wars: The Politics of Sex. Bloomsbury Publishing.

MacRoberts, M.H. & MacRoberts, B.R.. (1989). Problems of Citation Analysis: A Critical Review. *Journal of the American Society for Information Science*, 40(5): 342–349.

Maliniak, D., Powers, R. M., & Walter, B. F. (2013). The gender citation gap in international relations. *International Organization,* *67*(4), 889–922. https://doi.org/10.1017/S0020818313000209

Merton, R.K. (1968). The Matthew Effect in Science. *Science*, 159(3810): 56–63.

Merton, R. K. (1973). *The Sociology of Science: Theoretical and Empirical Investigations*. University of Chicago Press.

Mihaljević, H., Tullney, M., Santamaría, L., & Steinfeldt, C. (2019). Reflections on gender analyses of bibliographic corpora. *Frontiers in big Data*, *2*, 29.

Miller, D. I., Nolla, K. M., Eagly, A. H., & Uttal, D. H. (2018). The development of children's gender-science stereotypes: A meta-analysis of 5 decades of US draw-a-scientist studies. *Child development*, *89*(6), 1943-1955.

Milojević, S. (2010). Modes of collaboration in modern science: Beyond power laws and preferential attachment. *Journal of the American Society for Information Science and Technology, 61*(7), 1410–1423. https://doi.org/10.1002/asi.21331.

Moss-Racusin, C. A., Dovidio, J. F., Brescoll, V. L., Graham, M. J., & Handelsman, J. (2012). Science faculty's subtle gender biases favor male students. *Proceedings of the National Academy of Sciences, 109*(41), 16474–16479.

National Science Foundation. (2006). Science and Engineering Indicators. Chapter 5: Academic Research and Development. Data and Terminology. Retrieved from http://www.nsf.gov/statistics/seind06/c5/c5s3.htm#sb1.

Nick, I. M. (2017). Names, grades, and metamorphosis: A small-scale socio-onomastic investigation into the effects of ethnicity and gender-marked personal names on pedagogical assessments. *Names, 65*(3), 129–142.

Nosek, B. A., Smyth, F. L., Sriram, N., Lindner, N. M., Devos, T., Ayala, A., ... & Greenwald, A. G. (2009). National differences in gender–science stereotypes predict national sex differences in science and math achievement. *Proceedings of the National Academy of Sciences*, *106*(26), 10593-10597.

Oxfam (2016). An Economy for the 1%: How privilege and power in the economy drive extreme inequality and how this can be stopped. Oxfam Briefing Paper 210. Oxford: Oxfam International.

Pareto, V. (1896). *Cours d'économie politique*, Vol. 1. Lausanne : F. Rouge.

Peirce, C. S. (1933). Prolegomena to an apology for pragmaticism. In C. Hartshorne & P. Weiss (Eds.), *Collected papers of Charles Sanders Peirce: Vol. 4. The simplest mathematics* (para. 537). Harvard University Press. (Original work published 1906).

Price, D. J. D. S. (1976). A general theory of bibliometric and other cumulative advantage processes. *Journal of the American Society for Information Science*, *27*(5), 292–306.

Rajkó, A., Herendy, C., Goyanes, M., & Demeter, M. (2023). The Matilda effect in communication research: The effects of gender and geography on usage and citations across 11 countries. Communication Research. Advance online publication. https://doi.org/10.1177/00936502221124389.

Reisner, S. L., Conron, K. J., Baker, K., Herman, J. L., Lombardi, E., Greytak, E. A., Gill, A. M., Matthews, A. K., & Scout. (2015). "Counting" transgender and gender-nonconforming adults in health research: Recommendations from the gender identity in US surveillance group. Transgender Studies Quarterly, 2(1), 34–57.

Rossi, A.S. (1965). Naming Children in Middle-Class Families. *American Sociological Review*, 30(4): 499–513.

Rossiter, M. W. (1993). The Matthew Matilda effect in science. *Social Studies of Science, 23*(2), 325–341.

Santamaría, L., & Mihaljević, H. (2018). Comparison and benchmark of name-to-gender inference services. PeerJ Computer Science, 4, e156. https://doi.org/10.7717/peerj-cs.156.

Schelling, T. C. (1978). *Micromotives and Macrobehavior*. W. W. Norton.

Singh Chawla, D. (2019). Hyperauthorship: global projects spark surge in thousand-author papers. Nature, 576(7786), 186–187. https://doi.org/10.1038/d41586-019-03862-0.

Šolc, T. (2025). Unidecode: ASCII transliterations of Unicode text (Version 1.4.0) . PyPI. https://pypi.org/project/Unidecode/.

Sproat, R., & Shih, C. (1990). A statistical method for finding word boundaries in Chinese text. *Computer Processing of Chinese and Oriental Languages, 4*(4), 336–351.

Steinpreis, R. E., Anders, K. A., & Ritzke, D. (1999). The impact of gender on the review of curricula vitae of job applicants and tenure candidates: A national empirical study. *Sex Roles, 41*(7–8), 509–528.

Sue, C.A & Telles, E.E. (2007). Assimilation and Gender in Naming. *American Journal of Sociology*, 112 (5): 1383–1415.

Sugimoto, C. R., & Larivière, V. (2018). *Measuring Research: What Everyone Needs to Know*. Oxford University Press.

Sugimoto, C. R., & Larivière, V. (2023). *Equity for women in science: Dismantling systemic barriers to advancement*. Harvard University Press.

Teich, E. G., Kim, J. Z., Lynn, C. W., Simon, S. C., Klishin, A. A., Szymula, K. P., Srivastava, P., Bassett, L. C., Zurn, P., Dworkin, J. D., & Bassett, D. S. (2022). Citation inequity and gendered citation practices in contemporary physics. Nature Physics, 18(10), 1161–1170. https://doi.org/10.1038/s41567-022-01770-1.

Tscharntke, T., Hochberg, M. E., Rand, T. A., Resh, V. H., & Krauss, J. (2007). Author sequence and credit for contributions in multiauthored publications. *PLoS biology*, *5*(1), e18.

Torgrimson, B., & Minson, C. T. (2015). Sex and gender: What is the difference? *Journal of Applied Physiology, 99*(3), 785–787.

Traag, V. A., & Waltman, L. (2022). Causal foundations of bias, disparity and fairness. *arXiv*. https://arxiv.org/abs/2207.13665

Tregenza, T. (2002). Gender bias in the refereeing process? *Trends in Ecology & Evolution, 17*(8), 349–350.

Tripodi, F. (2023). Ms. categorized: Gender, notability, and inequality on Wikipedia. *New Media & Society, 25*(7), 1687–1707.

Tygert, M. (2021a). Cumulative deviation of a subpopulation from the full population. *Journal of Big Data*, 8(117), 1–60.

Tygert, M. (2021b). A graphical method of cumulative differences between two subpopulations. *Journal of Big Data*, *8*, 158. https://doi.org/10.1186/s40537-021-00540-9

Urbatsch, R. (2018). Feminine-sounding names and electoral performance. *Electoral Studies, 55*, 54–61.

Van den Besselaar, P., & Sandström, U. (2017). Vicious circles of gender bias, lower positions, and lower performance: Gender differences in scholarly productivity and impact. PLOS ONE, 12(8), e0183301. https://doi.org/10.1371/journal.pone.0183301

Van Fleet, D. D., & Atwater, L. (1997). Gender-neutral names: Don't be so sure! *Sex Roles, 37*, 111–123.

Vasilescu, B., Capiluppi, A., & Serebrenik, A. (2014). Gender, representation and online participation: A quantitative study. *Interacting with Computers, 26*(5), 488–511.

Walker, P. L., & Cook, D. C. (1998). Brief communication: Gender and sex: Vive la différence. *American Journal of Physical Anthropology, 106*(2), 255–259.

Webb, T. J., O'Hara, B., & Freckleton, R. P. (2008). Does double-blind review benefit female authors?. *Trends in Ecology & Evolution*, *23*(7), 351-353.

West, C., & Zimmerman, D. H. (1987). Doing gender. *Gender & Society, 1*(2), 125–151.

West, J. D., Jacquet, J., King, M. M., Correll, S. J., & Bergstrom, C. T. (2013). The role of gender in scholarly authorship. *PLOS ONE, 8*(7), e66212.

Westbrook, L., & Saperstein, A. (2015). New categories are not enough: Rethinking the measurement of sex and gender in social surveys. *Gender & Society, 29*(4), 534–560.

Wu, C. (2023). The gender citation gap: Why and how it matters. Canadian Review of Sociology, 60(2), 188–211. https://doi.org/10.1111/cars.12428

Wu, C. (2024). The gender citation gap: Approaches, explanations, and implications. Sociology Compass, 18(2), e13189. https://doi.org/10.1111/soc4.13189.

Wuchty, S., Jones, B. F., & Uzzi, B. (2007). The increasing dominance of teams in production of knowledge. *Science, 316*(5827), 1036–1039.

Xie, Y. & Shauman, KA. (2003). *Women in Science: Career Processes and Outcomes*. Harvard University Press.

Zheng, X., Yuan, H., & Ni, C. (2022). Gender gaps in academic career: The gender gap in academic career achievements and the mediation effect of work-family conflict and partner support. https://doi.org/10.1101/2022.03.23.485507

Zipf, G. K. (1949). *Human Behavior and the Principle of Least Effort*. Addison-Wesley.